\documentclass[sigconf,9pt]{acmart}
\usepackage[english]{babel}
\usepackage{graphicx}
\usepackage{multirow}
\usepackage{array}
\usepackage{tabularx}
\usepackage{pifont}
\usepackage{xcolor}

\usepackage{blindtext}

\usepackage{color,balance,xspace,verbatim,ifthen,engord}

\usepackage{gensymb,amsmath,wasysym,amsthm,marvosym}

\usepackage{booktabs,colortbl,diagbox,multirow,tabularx,tablefootnote} 

\usepackage{dblfloatfix} 
\usepackage{courier} 

\usepackage[ruled,vlined]{algorithm2e}
\usepackage{url}

\usepackage{tikz}

\usepackage{amsmath}
\usepackage{array}
\usepackage{url}
\usepackage{float}
\usepackage{graphicx}
\usepackage{multicol}
\usepackage{tabu}
\usepackage{listings}
\usepackage{makecell}
\usepackage{enumerate}
\usepackage{array}
\usepackage{color}

\usepackage{threeparttable}
\usepackage{graphicx}
\usepackage{subcaption}
\usepackage{booktabs}
\usepackage{xcolor}
\usepackage{appendix}
\usepackage{amsthm}
\usepackage{booktabs}
\usepackage{enumitem}
\usepackage{blindtext}
\usepackage{textcomp}
\usepackage{makecell}
\usepackage{multirow}
\usepackage{tabularx}
\usepackage[normalem]{ulem}

\newcommand{\COMMENTS}{yes}

\newcommand{\secref}[1]{\S\ref{#1}}
\newcommand{\figref}[1]{Figure~\ref{#1}}
\newcommand{\tabref}[1]{Table~\ref{#1}}

\newcommand{\algref}[1]{Algorithm~\ref{#1}}
\newcommand{\fig}[1]{Figure~\ref{#1}}

\ifthenelse{\equal{\COMMENTS}{yes}}{
  \newcommand{\todo}[1]{\textcolor{red}{\textbf{TODO:} #1}}
  \newcommand{\neil}[1]{\textcolor{blue}{\textbf{Nie:} #1}} 
  \newcommand{\grace}[1]{\textcolor{purple}{(grace: #1)}}
  \newcommand{\fye}[1]{\textcolor{red}{#1}}  
  \newcommand{\remind}[1]{\footnote{\textit{\textcolor{red}{\textbf{Remind:} #1}}}}
  
  \newcommand{\del}[1]{\color{blue} {\sout{#1}}}
  \newcommand{\p}[1]{\vskip 1ex \noindent\colorbox{yellow}{\parbox{\columnwidth}{#1}}\vskip 4pt}
  \newcommand{\note}[1]{\vskip 4ex \noindent\colorbox{yellow}{\parbox{\columnwidth}{#1}}\vskip 6ex} 
  \newcommand{\q}[1]{\vskip 1ex \noindent\colorbox{magenta}{\parbox{\columnwidth}{\textbf{Question:} #1}}\vskip 4pt} 
  \newcommand{\qa}[1]{\hl{\textbf{Answer:} #1}}
  
}{
  \newcommand{\todo}[1]{}
  
  \newcommand{\fye}[1]{}
  \newcommand{\remind}[1]{}

  \newcommand{\del}[1]{}
  \newcommand{\p}[1]{}
  \newcommand{\note}[1]{}

  \newcommand{\q}[1]{}
  \newcommand{\qa}[1]{}  
  \newcommand{\grace}[1]{}
  \newcommand{\neil}[1]{}
}

\newcommand{\circled}[1]{\raisebox{.5pt}{\textcircled{\raisebox{-.9pt} {#1}}}}
\newcommand{\para}[1]{\noindent\textbf{#1}}

\newenvironment{packeditemize}{\begin{list}{$\bullet$}{\setlength{\itemsep}{2pt}\addtolength{\labelwidth}{-6pt}\setlength{\leftmargin}{12pt}\setlength{\listparindent}{\parindent}\setlength{\parsep}{1pt}\setlength{\topsep}{2pt}}}{\end{list}}

\copyrightyear{2025}
\acmYear{2025}
\setcopyright{acmlicensed}\acmConference[SIGCOMM '25]{ACM SIGCOMM 2025 Conference}{September 8--11, 2025}{Coimbra, Portugal}
\acmBooktitle{ACM SIGCOMM 2025 Conference (SIGCOMM '25), September 8--11, 2025, Coimbra, Portugal}
\acmDOI{10.1145/3718958.3750468}
\acmISBN{979-8-4007-1524-2/2025/09}
\acmPrice{}

\newcommand{\SYS}{\textit{\textbf{InfiniteHBD}}\xspace}
\newcommand{\sys}{\textit{InfiniteHBD}\xspace}
\newcommand{\ocstrx}{\textit{OCSTrx}}
\newcommand{\revised}[1]{\textcolor{black}{#1}}
\newcommand{\docs}{\ocstrx}

\begin{document}


\title{InfiniteHBD: Building Datacenter-Scale High-Bandwidth Domain for LLM with Optical Circuit Switching Transceivers}

\renewcommand{\shorttitle}{\sys}

\author{Chenchen Shou$^{1,2,3}$ \hspace{0.5em} Guyue Liu$^{1,\dag}$ \hspace{0.5em} Hao Nie$^{2,\dag}$ \hspace{0.5em} Huaiyu Meng$^{3,\dag}$  \hspace{0.5em} Yu Zhou$^2$ \hspace{0.5em} \\ Yimin Jiang$^4$ \hspace{0.5em}  Wenqing Lv$^3$ \hspace{0.5em} Yelong Xu$^3$ \hspace{0.5em} Yuanwei Lu$^2$ \hspace{0.5em} Zhang Chen$^3$ \hspace{0.5em} \\ Yanbo Yu$^2$ \hspace{0.5em} Yichen Shen$^3$ \hspace{0.5em} Yibo Zhu$^2$ \hspace{0.5em} Daxin Jiang$^2$
}

\affiliation{
\institution{
$^1$Peking University \hspace{0.5em}
$^2$StepFun \hspace{0.5em}
$^3$Lightelligence Pte. Ltd. \hspace{0.5em}
$^4$Unaffiliated \hspace{0.5em}
}
\country{}
}


\renewcommand{\shortauthors}{Chenchen Shou et al.}

\begin{abstract}

Scaling Large Language Model (LLM) training relies on multi-dimensional parallelism, where High-Bandwidth Domains (HBDs) are critical for communication-intensive parallelism like Tensor Parallelism. However, existing HBD architectures face fundamental limitations in scalability, cost, and fault resiliency: switch-centric HBDs (e.g., NVL-72) incur prohibitive scaling costs, while GPU-centric HBDs (e.g., TPUv3/Dojo) suffer from severe fault propagation. Switch-GPU hybrid HBDs (e.g., TPUv4) take a middle-ground approach, but the fault explosion radius remains large.

We propose \textbf{\sys{}}, a transceiver-centric HBD architecture that integrates connectivity and dynamic switching at the transceiver level by embedding Optical Circuit Switching (OCS) within each transceiver.
It enables reconfigurable point-to-multipoint communication and scalable variable-size ring topologies. \sys{} achieves datacenter-scale scalability without cost explosion, fault isolation at the node level, and full bandwidth utilization for healthy GPUs.
Key innovations include a Silicon Photonic-based OCS transc\-eiver (\textbf{\ocstrx{}}), a reconfigurable k-hop ring topology, and an HBD-DCN orchestration algorithm.
The evaluation demonstrates that \sys{} reduces cost to \textbf{31\%} of NVL-72, achieves a \textbf{near-zero} GPU waste ratio (over 10x lower than NVL-72 and TPUv4), maintains \textbf{near-zero} cross-ToR traffic under 7\% node fault ratio, and improves Model FLOPs Utilization by \textbf{3.37×} compared to NVIDIA DGX (8 GPUs/node).

\end{abstract}



\renewcommand{\authors}{Chenchen Shou, Guyue Liu, Hao Nie, Huaiyu Meng, Yu Zhou, Yimin Jiang, Wenqing Lv, Yelong Xu, Yuanwei Lu, Zhang Chen, Yanbo Yu, Yichen Shen, Yibo Zhu, Daxin Jiang}

\maketitle

\renewcommand{\thefootnote}{\fnsymbol{footnote}}
\footnotetext[2]{Guyue Liu, Hao Nie, and Huaiyu Meng contributed equally to this work and share the corresponding authorship.}
\renewcommand{\thefootnote}{\arabic{footnote}}

\section{Introduction}

Large Language Models (LLMs) training relies on various parallelism strategies~\cite{megatrontrain3DP,zero}, such as Tensor Parallelism (TP), Expert Parallelism (EP), Data Parallelism (DP), Pipeline Parallelism (PP), Context Parallelism (CP) and Sequence Parallelism (SP). These strategies communicate over two types of AI datacenter compute fabrics, each with distinct bandwidth requirements. First, Datacenter Networks (DCNs) provide hundreds of Gbps per GPU and primarily handle DP, PP, CP, and SP traffic, which has lower communication demands. Second, High-Bandwidth Domains (HBDs) offer Tbps-level bandwidth, which is crucial for communication-intensive TP and EP. Efficient HBD design can reduce communication overhead, thereby improving Model FLOPs Utilization (MFU) - a key performance metric for LLM training.

The community has made significant advancements in designing DCNs for LLM training~\cite{wang2024railonly, sigcomm2024hpn,rail-optimized,sigcomm2024meta}. However, scaling HBD to optimize MFU in LLM training remains a challenging problem. Existing HBD architectures~\cite{nvl72,isca2023tpu,dojo,cacm2020tpuv3,aws-trainium} take important steps but still suffer from fundamental limitations in scalability, cost, and fault resiliency. 
\begin{packeditemize}
    \item \textbf{Switch-centric HBDs}, such as NVIDIA NVL-72~\cite{nvl72}, build multilayer \revised{non-blocking networks} for HBD with switch chips. However, the switch fabric incurs superlinear cost growth as it scales, constraining the number of GPUs per HBD. This limitation prevents optimal large TP and EP and causes severe \textit{resource fragmentation} when the size of TP/EP group increases. For instance, with 2 HBDs (32 GPUs each), 30 GPUs are wasted for TP-16 jobs if each HBD has a single GPU failure. This waste reduce to 14 GPUs if the two HBDs are combined into a 64-GPU unit.
    \item \textbf{GPU-centric HBDs}, such as Dojo~\cite{dojo}, NVIDIA V100~\cite{v100}, TPUv3~\cite{cacm2020tpuv3}, and SiP-Ring~\cite{sip-ml}, adopt low-cost GPU-to-GPU links to construct large-scale ring or mesh topologies, forwarding traffic directly through GPUs. However, these architectures suffer from a large \textit{fault explosion radius}, where a single GPU failure degrades bandwidth for a group of adjacent GPUs, compromising the entire topology. For example, in SiP-Ring, one single GPU failure breaks the ring and reforms the topology into a line. 
    \item \textbf{Switch-GPU Hybrid HBDs}, TPUv4~\cite{isca2023tpu} alleviates the limitation of a large fault explosion radius via OCS-based switches\footnote{In this paper, "OCS" specifically denotes \textit{optical circuit switching capability}, while OCS-based switch denotes \textit{optical circuit switch}.}: each set of 64 TPUs is connected as a cube, with these cubes connected to multiple OCS-based switches to isolate faults within the cubes. However, it does not fundamentally resolve the issue, as the fault explosion radius remains large at the cube level (64 TPUs).
\end{packeditemize}

In this paper, we take a first-principles approach to redesigning HBD for LLM training workloads.
Through a top-down analysis of parallelism strategies (\S\ref{sec:background:workload})  for maximizing MFU, we show that increasing TP size yields the most significant MFU gains for both dense and sparse LLM models. For large dense models~\cite{llama3herdmodels}, the optimal TP size scales from 16 to 64 as the number of GPUs increases. For sparse MoE models~\cite{hunyuanlarge,deepseekv3,mixtralexperts}, enlarging TP improves MFU more effectively than EP, particularly when considering the expert imbalance problem~\cite{sigcomm2023_janus}.

These findings lead to two key design principles: i) HBD should be optimized exclusively for TP Ring-Allreduce communication with large message sizes, which communicates with only logical neighboring nodes, eliminating the need for EP and the associated any-to-any communication; ii) Supporting large and adaptable TP is essential, as different GPU numbers and model sizes require varying TP configurations to maximize MFU.

Based on these principles, we propose \textbf{\sys}, a scalable and fault-resilient HBD architecture designed for optimizing TP communication. Our key insight is \textit{unifying connectivity and dynamic switching at the transceiver level} using OCS. By embedding OCS in each transceiver, we achieve reconfigurable point-to-multipoint connectivity. This marks a departure from traditional designs, where transceivers support only point-to-point connections and rely on high-radix switches for routing. We call this new design \textbf{transceiver-centric HBD architecture}. This transceiver-centric architecture offers two key benefits: i) It enables the flexible construction of arbitrarily large ring topologies by intra-node loopback mechanism. This can support optimal TP group sizes for different models, while effectively minimizing resource fragmentation; ii) When one node fails, its neighboring transceivers dynamically reconfigure connections to reroute traffic, significantly reducing the fault explosion radius and improving system resilience.

We realize the transceiver-centric HBD architecture in production by combining the following key ideas: 
\begin{packeditemize}
    \item \emph{Silicon Photonics based OCS transceiver (OCSTrx):} To design a cost-effective low-power transceiver with OCS support, we leverage the current advances of Silicon Photonics (SiPh) technology. Compared to MEMS~\cite{missionapollo, mem-optical-switches} technology which has been widely used to realize OCS, SiPh offers simpler structures, lower cost and power consumption. We build OCS with Mach-Zehnder interferometer (MZI) matrix~\cite{mzi}, taped out with $65nm$ CMOS processes. The chip area is smaller than $136.5mm^2$ while the chip power consumption is $3.2Watts$, which can be integrated into commercial QSFP-DD $800Gbps$ transceiver~\cite{QSFP-DD} with 60-80 $\mu s$ reconfiguration latency. \revised{\ocstrx{} introduces zero additional bit error rate in most cases, with an average insertion loss of 3.3dB at room temperature.}
    \item \emph{Reconfigurable K-Hop Ring Topology:} While OCSTrx offers reconfigurable connections at the transceiver level, constructing adaptive-size rings that maximize GPU utilization remains a challenge. For example, a naive full-mesh topology built with OCSTrx would impose strict limits on TP size, while also resulting in significant bandwidth waste and fragmentation. To address this, we propose a reconfigurable K-Hop Ring, where each node connects to all other nodes within $\le K$ hops via \textit{OCSTrx}. The intra-node \textit{loopback} mechanism enables dynamic ring construction, while the inter-node \textit{backup link} bypasses faulty nodes, ensuring high fault tolerance.
    
    \item \emph{HBD-DCN Orchestration Algorithm:} While an optimal HBD topology is critical, end-to-end training performance also depends on efficient HBD-DCN coordination. For example, the orchestration of TP groups in HBD directly determines DP traffic distribution, which impacts congestion in DCN, ultimately governing training performance. Unfortunately, existing approaches lack mechanisms to jointly coordinate DCN and HBD to mitigate congestion and optimize communication efficiency. To address this, we propose a new orchestration algorithm that minimizes cross-ToR traffic, thereby minimizing congested traffic.
\end{packeditemize}

To the best of our knowledge, \sys is the first HBD design capable of scaling to datacenter scale while avoiding cost explosion and increased failure-induced waste. We evaluated \sys \xspace with the real 348-day fault trace from our 3K-GPU cluster\footnote{\revised{Details in Appendix~\S\ref{appendix:production-fault-trace}. We have open-sourced this trace at \url{https://github.com/stepfun-ai/InfiniteHBD-Trace}.}}. When executing TP32 jobs with the trace, it demonstrates 0.53\% GPU waste ratio - 20x and 14x lower than NVL-72 (10.04\%) and TPUv4 (7.56\%). It achieves 3.24x and 1.59x cost reductions compared to NVIDIA NVL-72 and Google TPUv4 respectively. Through the orchestration algorithm, it maintains near-zero cross-ToR traffic under 7\% node failure rates. Its dynamic ring formation capability enables 3.37x higher MFU than NVIDIA DGX systems~\cite{dgx} (8 GPUs/node). 

This work does not raise any ethical issues.

\section{Background and Motivation}
\label{sec:background}
In this section, we first introduce LLM training in AI datacenters (DCs) (\S\ref{sec:background:llm_training}). Then, we examine existing High-Bandwidth Domain (HBD) architectures and discuss their limitations (\S\ref{sec:background:hbd}). Finally, we summarize key design principles of HBD for LLM training (\S\ref{sec:background:workload}).

\subsection{LLM Training in AI DC}
\label{sec:background:llm_training}

\begin{figure*}[!t]
\centering
\begin{subfigure}[b]{0.25\textwidth}
    \centering
    \includegraphics[height=16ex]{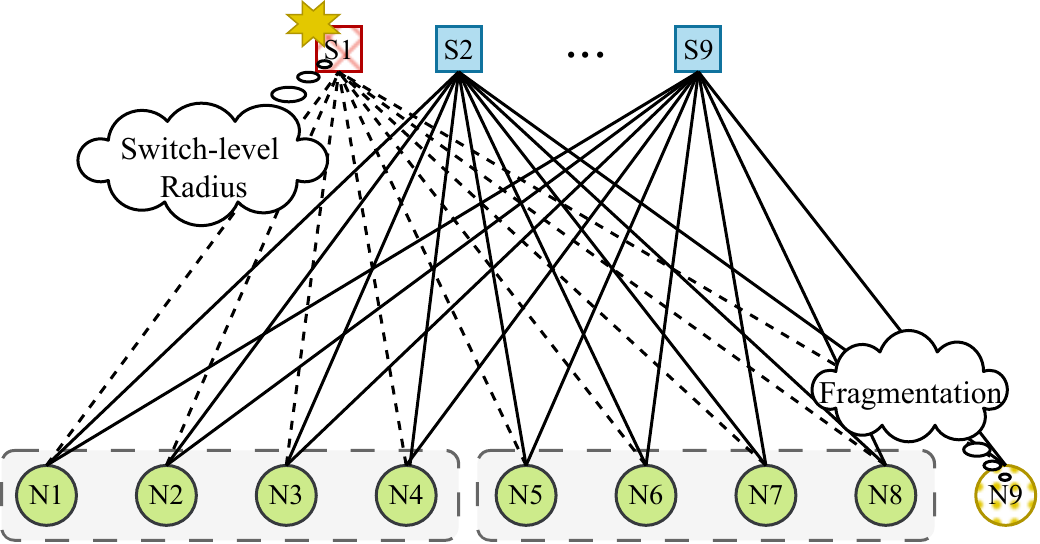}
    \caption{Switch-centric: NVL36}
    \label{fig:hbd-archs:nvl36}
\end{subfigure}
\hspace{-1ex}\hfil\hspace{-1ex}
\begin{subfigure}[b]{0.24\textwidth}
    \centering
    \includegraphics[height=16ex]{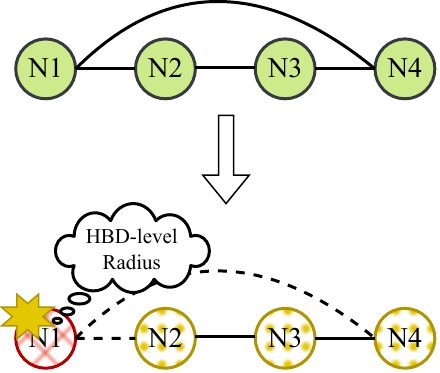}
    \caption{GPU-centric: SiP-Ring}
    \label{fig:hbd-archs:sip-ring}
\end{subfigure}
\hspace{-1ex}\hfil\hspace{-1ex}
\begin{subfigure}[b]{0.24\textwidth}
    \centering
    \includegraphics[height=16ex]{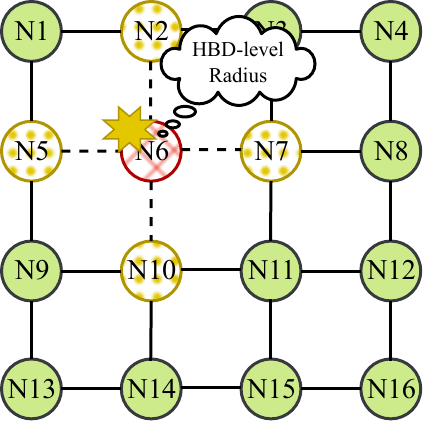}
    \caption{GPU-centric: Dojo}
    \label{fig:hbd-archs:dojo}
\end{subfigure}
\hspace{-1ex}\hfil\hspace{-1ex}
\begin{subfigure}[b]{0.25\textwidth}
    \centering
    \includegraphics[height=16ex]{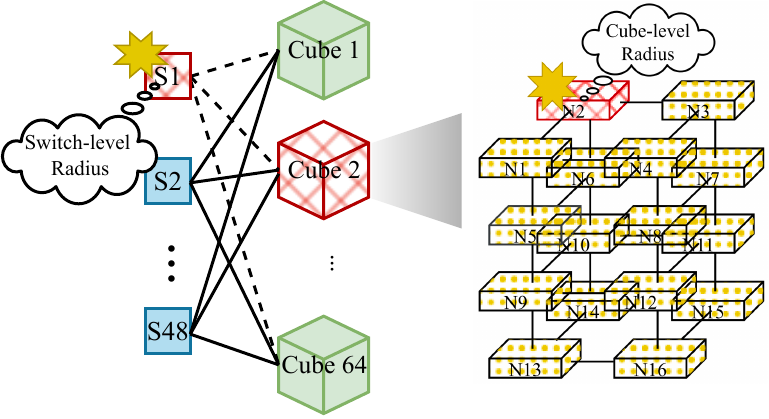}
    \caption{Hybrid: TPUv4}
    \label{fig:hbd-archs:tpuv4}
\end{subfigure}
\vspace{-2ex}
\caption{Illustrative examples of HBD architectures. N represents Node, and S represents Switch. Red (with cross hatch) represents fault device and yellow (with dots) represents unavailable or downgraded GPU.}
\label{fig:hbd-archs}
\vspace{-1ex}
\end{figure*}

\begin{table*}[!htbp]\small
\centering
\begin{tabular}{llllllll}
\toprule
\multirow{2}{*}{\textbf{Architecture}}  & \multirow{2}{*}{\textbf{Type}}  & \multirow{2}{*}{\textbf{Scalability}} & \multirow{2}{*}{\begin{tabular}[c]{@{}l@{}}\textbf{Collective} \\ \textbf{Primitives}\end{tabular}} & \multicolumn{2}{l}{\textbf{Fault Explosion Radius}} & \multirow{2}{*}{\begin{tabular}[c]{@{}l@{}}\textbf{Interconnect} \\ \textbf{Cost}\end{tabular}} & \multirow{2}{*}{\textbf{Fragmentation}} \\
                              &                                                &                              &                                                                                   & \textbf{Node-Side}           & \textbf{Switch-Side}          &                                                                               &                                \\
\midrule
NVL                           & Switch-centric                       & Low                          & Full CCL                                                                          & Node-level          & Switch-level         & High                                                                          & Many                           \\
\makecell{Dojo, TPUv3, SiP-Ring}         & GPU-centric                       & High                         & Ring-Allreduce                                                                    & HBD-level           & \ding{55}            & Low                                                                           & Few                            \\
TPUv4, TPUv5p                         & Switch-GPU Hybrid                   & Moderate                     & Ring-Allreduce                                                                    & Cube-level          & Switch-level         & Moderate                                                                      & Few                            \\
\sys{}                   & Transceiver-centric & High                         & Ring-Allreduce                                                                    & Node-level          & \ding{55}            & Low                                                                           & Few  \\
\bottomrule
\end{tabular}
\caption{Comparative analysis of HBD architectures.}
\label{tab:hbd-compare}
\vspace{-6ex}
\end{table*}

\para{LLM training parallelism and communication.} LLM training jobs employ various parallelism strategies to efficiently utilize GPUs distributed across AI DCs~\cite{megatron-lm, zero}. Based on communication loads, parallelism can be categorized into two types. The first type is \textit{communication-intensive  parallelism} which involves high communication load. Tensor Parallelism (TP) splits the model across multiple GPUs and synchronizes via AllReduce. The ring algorithm for AllReduce is theoretically optimal~\cite{patarasuk2009bandwidth}, making ring-based topologies ideal for TP. Expert Parallelism (EP), designed for Mixture of Experts (MoE) models~\cite{hunyuanlarge,deepseekv3,mixtralexperts}, assigns experts to different GPUs and relies on AlltoAll communication, requiring topologies with high bisection bandwidth (e.g., Full-Mesh). In contrast, parallelism strategies such as Data Parallelism (DP), Pipeline Parallelism (PP), Context Parallelism (CP), and Sequence Parallelism (SP) introduce lower communication overhead, placing less  demands on network performance.

\para{Compute fabric. } Compute fabric in AI DC interconnects GPUs to efficiently transmit model gradients and parameters. It consists of two primary components: Datacenter Network (DCN) and High-Bandwidth Domain (HBD). 
DCN provides communication across the entire AI DC via Ethernet or Infiniband, the bandwidth is around $200\sim 800Gbps$. Widely used DCN architectures include Fat-Tree~\cite{sigcomm2008fattree} and Rail-Optimized~\cite{rail-optimized}. In comparison, HBD offers Tbps-level throughput, and is more suitable for TP/EP. However, its scale is typically constrained by interconnection costs and fault tolerance considerations. For example, NVL-72~\cite{nvl72} only interconnects 72 GPUs per HBD.

\para{Faults and fault explosion radius. }As revealed by current advances of AI DCs~\cite{sigcomm2024hpn, sigcomm2024meta}, training jobs experience a variety of faults, such as GPU faults, optical transceiver faults, switch faults, and link faults. We quantify the fault impact using the \textit{fault explosion radius}, defined as \textit{the number of GPUs degraded by a single fault event}.
The fault explosion radius varies depending on both the system architecture and the fault component.
For example, if a switch fails, the bandwidth of all devices connected to it will degrade, illustrating the switch-level fault explosion radius.

\para{HBD fragmentation.} When the number of GPUs in the HBD cannot be evenly divided by the size of the parallel group (i.e., TP size), the remaining GPUs become unusable, leading to resource waste.
The GPU waste ratio for each HBD can be expressed by the formula $\{(HBD_{size} - N_{fault}) \mod TP_{size}\}/{HBD_{size}}$.
In AI DCs with small-scale HBDs, GPU waste due to fragmentation is significant because each HBD experiences independent fragmentation.
This issue worsens as the TP group size increases with model scale. For example, for NVL-36 shown in \figref{fig:hbd-archs:nvl36}, running TP-16 causes $\geq$11\% GPU waste ratio.

\subsection{Limitations of Existing HBDs} 
\label{sec:background:hbd}

Existing HBD architectures for LLM training can be categorized into three types, based on the key components that provide connectivity. A summary is shown in Table~\ref{tab:hbd-compare}.

\para{Switch-centric HBD.}
This type of architecture leverages switch chips to interconnect GPUs, as shown in \figref{fig:hbd-archs:nvl36}.
A prominent example is NVIDIA, which utilizes NVLink and NVLink Switch ~\cite{nvlink,nvswitch}, e.g., DGX H100~\cite{dgx} with 8-GPU and GB200 NVL-36, NVL-72, and NVL-576~\cite{nvl72}. 
These architectures offer high-performance any-to-any communication.
However, switch-centric HBDs have several drawbacks: i) They require a large number of switch chips due to their limited per-chip throughput; ii) They are vulnerable to a switch-level fault explosion radius—when a switch chip fails, all connected nodes experience bandwidth degradation; iii) High interconnect costs constrain the scale of HBDs, leading to significant fragmentation when serving large models.

\para{GPU-centric HBD.}
GPU-centric HBD architectures construct the HBD using direct GPU-to-GPU connections, eliminating the need for switch chips. As a result, cost scales linearly with HBD size.
A representative example is SiP-Ring~\cite{sip-ml}, shown in \figref{fig:hbd-archs:sip-ring}, where GPUs are organized into fixed-size rings. However, this design imposes a strict limitation: the TP group size must remain fixed. 
To enable communication at dynamic scales and support a wider range of workloads, more complex topologies are adopted (e.g., Dojo~\cite{dojo}, NVIDIA V100~\cite{v100},  TPUv3~\cite{cacm2020tpuv3}, and AWS Trainium ~\cite{aws-trainium} ), which support dynamic scaling by allowing jobs to execute on topology subsets of varying sizes. As shown in \figref{fig:hbd-archs:dojo}, Dojo~\cite{dojo} connects GPUs via mesh-like topologies and employs GPUs to forward traffic. While GPU-centric architectures mitigate cost explosion and can support various scales, they suffer from a large fault explosion radius. A single GPU failure can disrupt the entire HBD by altering its connectivity, degrading communication performance even for healthy GPUs—such as the yellow GPUs in \figref{fig:hbd-archs:dojo}.

\para{Switch-GPU Hybrid HBD.}
This architecture interconnects GPUs via a combination of direct GPU-to-GPU connections and switch links. A typical example is TPUv4~\cite{isca2023tpu}, which organizes TPUs into $4^3$ TPU cubes and connects them via centralized OCS-based switches (\figref{fig:hbd-archs:tpuv4}). TPUv4 scales up to 4,096 TPUs, with its expansion primarily limited by the port count of the OCS-based switch. Furthermore, it suffers from a cube-level fault explosion radius—a failure in any single TPU affects the entire 64-TPU cube, leading to significant performance degradation. Furthermore, OCS-based switches face challenges of high costs and manufacturing complexity, which undermines the cost-effectiveness of TPUv4. TPUv5p cluster~\cite{tpuv5} is similar to TPUv4 but can scale out to 8,960 TPUs.

\begin{figure*}[!tp]
    \centering
    \includegraphics[width=\linewidth]{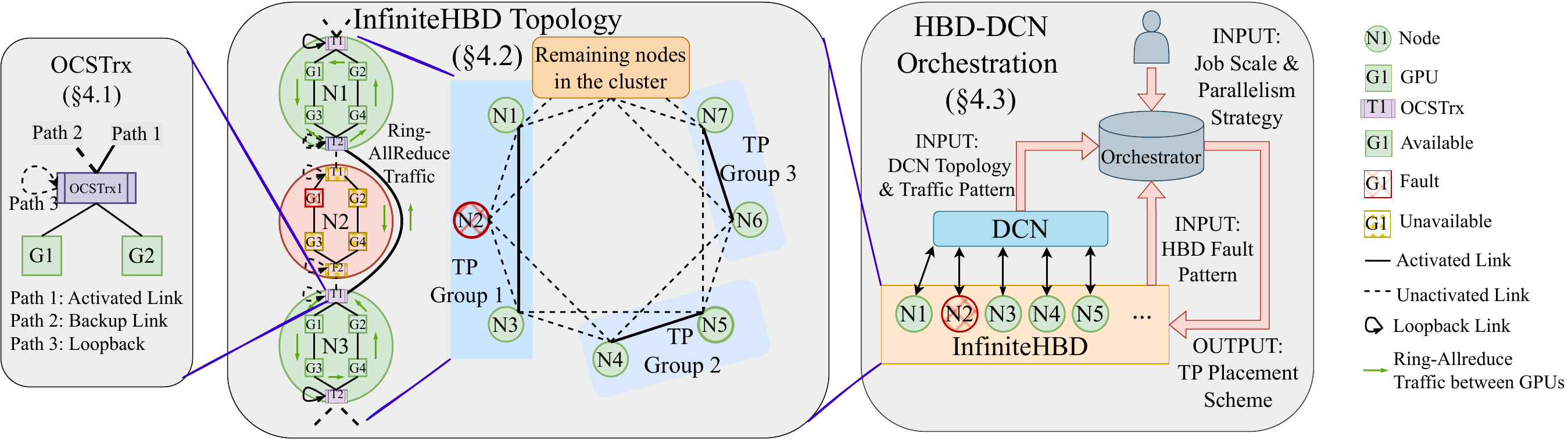}
    \vspace{-5ex}
    \caption{\sys{} overview.}
    \label{fig:overview}
    \vspace{-2ex}
\end{figure*}

\subsection{Key Attributes of An Ideal HBD}
\label{sec:background:workload}

\vspace{-1em}
\begin{table}[h!t] \small
    \centering
    \begin{tabular}{cllllll}
    \toprule
    \textbf{GPU} & \textbf{TP} & \textbf{PP} & \textbf{DP} & \textbf{MFU} & \textbf{$\textbf{MFU}_{TP-8}$} & \textbf{Improve}\\
    \midrule
    1024    & 16 & 4  & 16  & 0.5236 & 0.5217   & 1.0036      \\
    4096    & 16 & 8  & 32  & 0.4668 & 0.4282   & 1.0901      \\
    8192    & 32 & 8  & 32  & 0.4247 & 0.3512   & 1.2093      \\
    16384   & 32 & 16 & 32  & 0.3756 & 0.2584   & 1.4536      \\
    32768   & 32 & 16 & 64  & 0.3090 & 0.1690   & 1.8284      \\
    65536   & 64 & 16 & 64  & 0.2493 & 0.0999   & 2.4955      \\
    131072  & 64 & 16 & 128 & 0.1851 & 0.0550   & 3.3655      \\
    \bottomrule
    \end{tabular}
    \caption{Optimal parallelism strategy for maximum MFU of Llama 3.1-405b, compared to the baseline MFU for TP-8 (e.g., in widely-deployed NVLink architectures), when GPU number varies.}
    \label{tab:eval:llama3-optimal}
    \vspace{-2em}
\end{table}

Existing HBD architectures face fundamental limitations in interconnection cost, resource utilization, and failure resiliency when scaling. To guide a better design, we analyze existing training workloads and explore two key questions without the limitations imposed by current HBD: i) What is the optimal group size that HBD should support? ii) What traffic patterns should HBD accommodate?

\para{Large and adaptable TP size is critical for dense models.}
The optimal LLM training parallelism depends on model architectures and cluster configurations. For example, as illustrated by previous work~\cite{disttrain, nsdi2025_rlhfuse}. 
We evaluate the Model FLOPs Utilization (MFU) for Llama 3.1-405B~\cite{llama3.1-405b} using our in-house LLM training simulator (\S\ref{sec:simulation:end2end}) and report the results in Table~\ref{tab:eval:llama3-optimal}. MFU and TP/PP/DP columns denote the optimal MFU when TP size is unconstrained and the corresponding parallelism strategies respectively. $MFU_{TP-8}$ column denotes the optimal MFU when TP size is limited to 8. As we increase the number of GPUs, the optimal TP size grows from 16 to 64, a trend we observe across other large dense models. 
In this case, the HBD scale restricts the maximum size of TP, which affects training performance as a result. 

\begin{table}[h!t] \small
\vspace{-2ex}
\centering
\begin{tabular}{cccc}
\toprule
\multicolumn{2}{c}{\textbf{Parallelism}} & \textbf{Operation}  & \textbf{Traffic Load}  \\
\midrule

\multicolumn{2}{c}{TP}            & AllReduce     &$2bsh\cdot\frac{n-1}{n}$ \\ 
\multicolumn{2}{c}{EP}            & AllToAll     &$2bsh\cdot\frac{n-1}{n}\cdot\frac{k}{n}$\\
\bottomrule
\end{tabular}
\caption{Communication load of TP and EP on a single MoE layer. $b$: batch size; $s$: sequence length; $h$: hidden dim; $k$: topK of MoE router; $n$: parallel size. Assume each expert is assigned equal number of tokens.}
\label{tab:workload}
\vspace{-5ex}
\end{table}

\para{MoE can also be efficient with large-size TP.}
Beyond widely used dense models, we also examine sparse MoE models, which are trending toward larger scales (e.g., 1T parameters~\cite{switch_transformer}). The distributed training for MoE can be achieved through TP or EP (or a combination of them)\footnote{For TP, each expert is equally sharded to GPUs. For EP, each expert is indivisible and allocated to one GPU in the EP group.}~\cite{sigcomm2023_janus}, both TP and EP are communication-intensive~\cite{atc2023_lina}, making them heavily reliant on HBD.

\begin{table}[h!t]
    \vspace{-1ex}
    \centering
    \begin{tabular}{cccccc}
        \toprule
        & TP & \multicolumn{4}{c}{EP} \\
        \hline
        imbalance coef & - & 0\% & 10\% & 20\% & 30\% \\
        MFU (\%) & 31.2 & 31.5 & 30.5 & 29.8 & 28.8 \\
        \bottomrule
    \end{tabular}
    \caption{Performance comparison of TP and EP when training GPT-MoE.}
    \label{tab:ep-imbalance}
    \vspace{-2em}
\end{table}

Our production training experience on a 1T MoE model in production brings the following insights into the pros and cons of TP and EP.
On the one hand, EP is more communication-efficient than TP. Table~\ref{tab:workload} compares the communication volume of TP and EP. Clearly, EP is better if $k<n$, which is common~\cite{deepseekv3} because existing models often choose small $k$ for higher computation sparsity.
On the other hand, EP suffers from the well-known expert imbalance problem~\cite{sigcomm2023_janus}, especially when the MoE routers use the no-token-left-behind algorithm~\cite{deepseekv3, megablocks, glam}. This will result in a non-equivalent number of tokens that each expert will receive, which hence causes straggler nodes that waste GPU cycles of other nodes. 
\tabref{tab:ep-imbalance} shows the simulated result of training GPT-MoE with 1.1T parameters (details in Appendix~\S\ref{appendix:gpt-moe}) under different expert imbalance coefficients\footnote{Calculated as $\frac{max - min}{max}$, where $max$ and $min$ represent the maximum and minimum tokens allocated to each expert respectively.}. When $coef=0$, EP is better than TP due to smaller communication overhead. As $coef$ increases, the MFU drops because of the straggler issue.

\para{Key findings}. These experiments provide us with two key findings for HBD design. First, larger HBD size is increasingly needed for rapidly scaling LLMs (i.e., more than 1T parameters). Second, with larger HBD enabled, using TP is more favorable than EP to train an MoE model, because TP shards the computation equally across GPUs and hence bypasses the expert imbalance problem. 

These findings reveal two key design principles for HBD:  i) HBD must inherently support large and adaptable TP sizes, which fundamentally requires the scalability of HBD architecture; ii) the HBD designs need to ensure the effective support for the Ring-AllReduce communication. Given the demonstrated efficiency of TP in MoE training, ensuring support for Ring-AllReduce is sufficient for mainstream LLM training scenarios; iii) small fault explosion radius. Thus, \textit{\textbf{we propose designing a large and adaptable HBD architecture tailored for ring-based TP communication to optimize LLM parallelism strategies.}}

\section{Design Overview}
\label{sec:overview}

In this section, we first present our new HBD architecture \sys{} guided by the design principles outlined above. We then provide an overview of its key components.

\begin{figure*}[ht]
    \centering
    \begin{subfigure}[b]{0.45\textwidth}
        \centering
        \includegraphics[height=80pt]{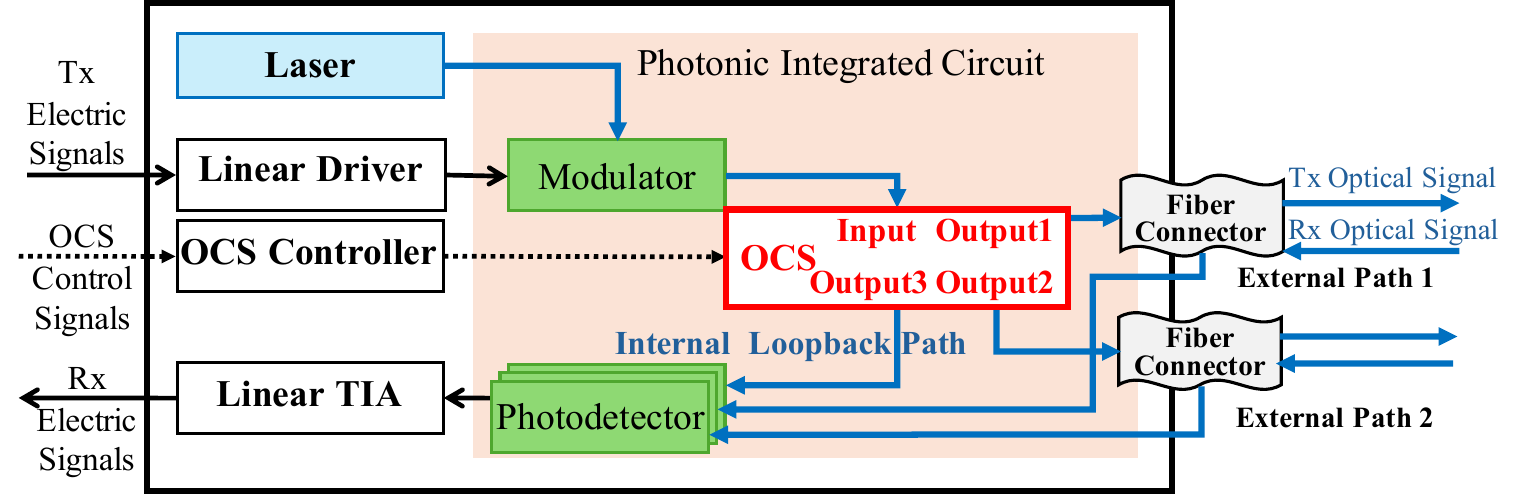}
        \caption{Components of OCS transceivers.}
        \label{figure:design:transceiver:component}
    \end{subfigure}
    \hspace{10pt}
    \begin{subfigure}[b]{0.45\textwidth}
        \centering
        \includegraphics[height=80pt]{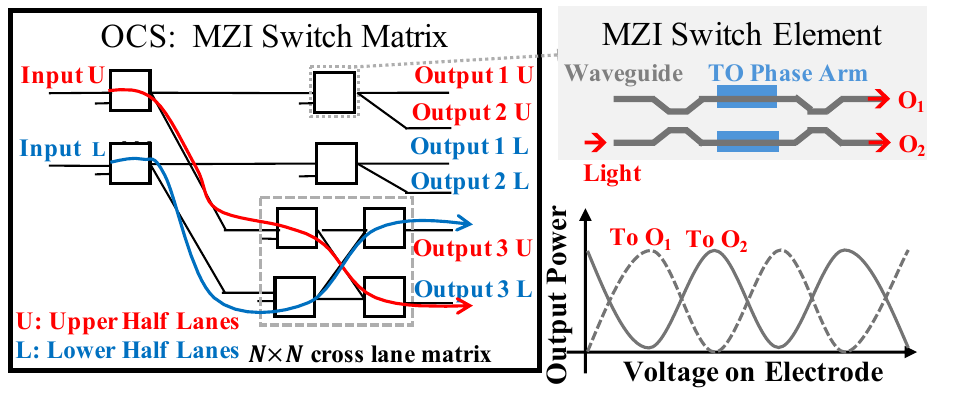}
        \caption{Zoom into OCS MZI switch matrix.}
        \label{figure:design:transceiver:ocs}
    \end{subfigure}
    \vspace{-10pt}
    \caption{Design of OCS Transceivers. The core component is OCS integrated in transceivers.}
    \label{figure:design:transceiver}
    \vspace{-15pt}
\end{figure*}

\para{Transceiver-centric HBD architecture}.
As discussed in \S\ref{sec:background:hbd} and summarized in \tabref{tab:hbd-compare}, existing architectures face a fundamental tradeoff among scalability, cost, and fault isolation. The GPU-centric architecture offers high scalability and low cost connectivity but suffers from a large fault explosion radius. In contrast, the switch-centric architecture improves fault isolation by leveraging centralized switches to confine failures to the node level. However, this comes at the cost of reduced scalability and higher connection overhead. The GPU-switch hybrid architecture takes a middle-ground approach but still suffers from significant fault propagation. As a result, no existing architecture fully meets all requirements.

Our key insight is that \textit{connectivity and dynamic switching can be unified at the transceiver level} using Optical Circuit Switching (OCS). By embedding OCS within each transceiver, we can enable reconfigurable point-to-multipoint connectivity, effectively combining both connectivity and switching at the optical layer. This represents a fundamental departure from conventional designs, where transceivers are limited to static point-to-point links and rely on high-radix switches for dynamic switching. We refer to this novel design as the \textit{transceiver-centric HBD architecture}. 

We realize this design with \sys{}, which has three key components as shown in \figref{fig:overview}.

\para{Design 1: Silicon Photonics-based OCS transceiver (OCSTrx) (\secref{sec:design:docs}).} 
To enable large-scale deployment, we require a low-cost, low-power transceiver with Optical Circuit Switching (OCS) support. Unlike prior high-radix switches solutions that rely on MEMS-based switching~\cite{missionapollo, mem-optical-switches}, we leverage advances in Silicon Photonics (SiPh), which offer a simpler structure, lower cost, and reduced power consumption—making them well-suited for commercial transceivers.

Our SiPh-based OCS transceiver (OCSTrx), shown on the left side of \figref{fig:overview}, provides two types of communication paths: i) \textit{Cross-lane loopback path (path 3)}, enabling direct GPU-to-GPU communication within the node, which can be used to construct dynamic size topologies; ii) \textit{Dual external paths (path 1\&2)}, connecting to external nodes. All these paths utilize time-division bandwidth allocation, featuring 60-80 $\mu s$ reconfiguration latency. With this capability, our \ocstrx{} allows the dynamic reallocation of full GPU bandwidth to a single active external path, rather than splitting it across multiple paths. This eliminates redundant link waste—for instance, activating one external path completely disables the other, ensuring efficient bandwidth utilization.

\para{Design 2: Reconfigurable K-Hop Ring topology (\secref{section:design:topology}).}
With \ocstrx~ that provides reconfigurable connections at the transceiver, the next challenge is designing the topology. A naive starting point is the full-mesh topology~\cite{fullmesh} which can provide full connectivity among all nodes using \ocstrx~. However, full-mesh design requires $O(N^2)$ links, inducing prohibitive complexity and cost. To reduce costs while maintaining near-ideal fault tolerance and performance, we prune the full-mesh topology into a K-Hop Ring topology based on traffic locality and fault non-locality (Details in~\S\ref{section:design:topology}).
\revised{The topology’s cost and wiring complexity grow linearly with node count, enabling high scalability.}
Combining the reconfigurability of \ocstrx{}, we propose a \textit{reconfigurable K-Hop Ring topology}, shown in the middle of \figref{fig:overview}, which consists of two key parts:

i) \textit{Intra-node topology:} dynamic GPU-granular ring construction is enabled by activating loopback paths. For example, while $N_1$-$N_3$ physically form a line topology, activating loopback paths creates a ring between $N_1$'s GPUs (1–4) and $N_3$'s GPUs (1–4). This mechanism allows for the construction of arbitrary-sized rings at any location, supporting optimal TP group sizes for different models while effectively minimizing resource fragmentation.

ii) \textit{Inter-node fault isolation: } dual external paths connect to primary and secondary neighbors (e.g., 2-Hop Ring). When a node fails (e.g., $N_2$), its neighbor ($N_1$) activates the backup path ($N_1$-$N_3$) to bypass the fault while maintaining full bandwidth, thus reducing the fault explosion radius to the node-level.
\S\ref{section:design:topology} generalizes this design to $K>2$.

\para{Design 3: HBD-DCN Orchestration Algorithm (\secref{sec:design:orch}).}
Designing an optimal HBD topology is crucial, but end-to-end training performance also depends on the efficient coordination between HBD and DCN. For instance, improper orchestration of TP groups can cause DP traffic to span across ToRs, resulting in DCN congestion. However, existing methods lack the ability to jointly optimize HBD and DCN coordination to alleviate congestion and enhance communication efficiency.
To address this, we propose the HBD-DCN Orchestration Algorithm, as shown on the right side of \figref{fig:overview}. The orchestrator takes three inputs: the user-defined job scale and parallelism strategy, the DCN topology and traffic pattern, and the real-time HBD fault pattern. It then generates the TP placement scheme, which maximizes GPU utilization and minimizes cross-ToR communication within the DCN.

\section{\SYS{} Design}
\label{section:design}

This section first introduces the innovative design of OCS transceivers (OCSTrx) based on Silicon Photonics (SiPh) chips (\secref{sec:design:docs}), a key enabler for \sys{}, providing both cost efficiency and reconfigurability. Next, we present the DC-scale \sys{} topology design (\secref{section:design:topology}) based on OCSTrx. Finally, we outline the HBD-DCN orchestration algorithm (\secref{sec:design:orch}), designed to optimize communication efficiency for training jobs.

\subsection{SiPh-based OCS transceiver (OCSTrx)}
\label{sec:design:docs}

The \ocstrx \xspace is designed for reconfigurable point-to-multipoint connectivity. It incorporates a compact OCS-based switch with three Rx/Tx paths, utilizing the MZI switch ~\cite{mzi} micro-structure with thermo-optic (TO) effect~\cite{thermo-optic_2006} phase arms. This OCS-based switch is seamlessly integrated into the Photonic Integrated Chip (PIC) of the transceiver, serving as the MZI switch matrix within the Tx light path, and providing photodetector (PD) modules for each Rx path.

\para{SiPh-Based OCS.} 
Currently, there are two predominant technological approaches for OCS.
Micro Electromechanical systems (MEMS)~\cite{missionapollo, mem-optical-switches} are attractive for commercial adoption because they support large port radix, up to a $320\times 320$ matrix~\cite{mems-320}.
Another option is SiPh-based OCS. 
Its structure is simpler and cheaper to manufacture, the limitation is its radix due to optical losses in the multistage light path selector. 
Given that the locality of traffic and external paths count of \ocstrx{} is only two, SiPh-based OCS offers greater advantages. 

So we choose the design of an SiPh-based OCS using the MZI micro-structure~\cite{mzi}.
The basic mechanism of MZI switch elements is controlling the phase difference between light paths in two phase arms, and then directs the output light to specific ports through interference at the output combiner. TO effect is utilized for phase arm control, for lower reconfiguration latency compared to MEMS.

\para{OCS Micro-Structure Design.} As shown in Figure~\ref{figure:design:transceiver:component}, the initial routing decision is made by two MZI switch elements, determining whether to direct the signal through external output 1\&2, or the internal loopback path.
Subsequently, an internal $N\times N$ MZI switch matrix is incorporated to facilitate the cross-lane loopback mechanism, exemplified by the blue and red paths.  Notably, this design can reduce stages count and light attenuation of output 1\&2, while ensuring consistent light attenuation for them. The design is implemented on the Photonic Integrated Circuit (PIC) chip.

\para{Transceiver Design.} In \ocstrx, the Tx electrical signal is amplified by linear driver and converted to optical signal by modulators as in \figref{figure:design:transceiver}. One laser is coupled into the PIC as the optical source.
On the receiving end, multiple photodetectors capture the Rx optical signal from all available paths separately. The output from the activated photodetector is then amplified by a linear transimpedance amplifier (TIA).
\ocstrx{} offers significant benefits, including high compactness, low power consumption, and cost-effective mass production.
\revised{Moreover, \ocstrx{} is integrated into commercial transceivers, and its failures manifest as regular transceiver failures without introducing new failure patterns.}

\subsection{\sys Topology}
\label{section:design:topology}

In this section, we present the \sys{} topology design (\figref{fig:overview}) integrating \docs{} that allows all GPUs within the datacenter to be connected in a \textit{reconfigurable K-Hop Ring topology}, while supporting dynamic ring construction and high fault tolerance.  

\begin{figure}[h!t]
   \centering
   \includegraphics[width=0.8\linewidth]{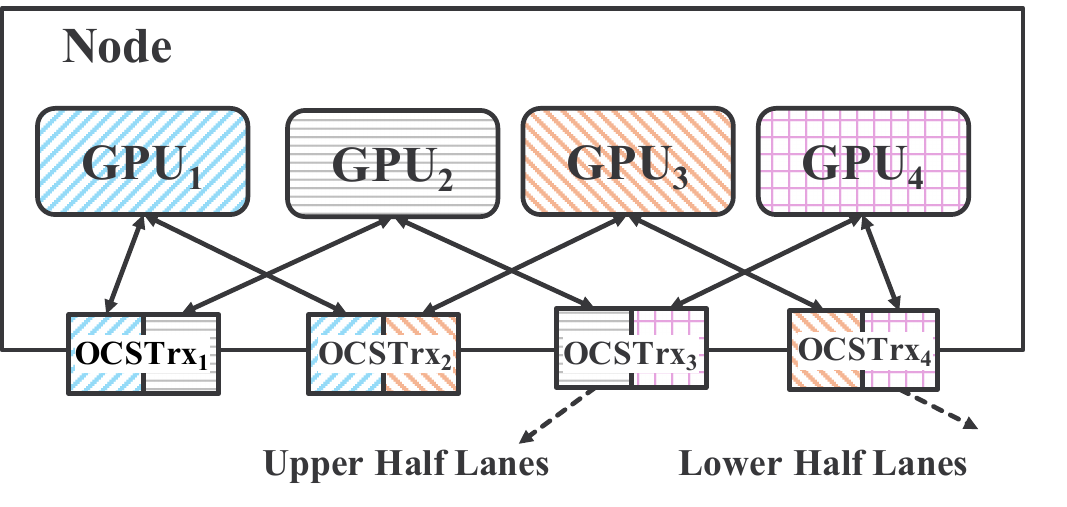}
   \vspace{-2ex}
   \caption{\ocstrx{} connection within nodes. Each block contains multiple \ocstrx \xspace as one bundle, e.g., $8\times 800Gbps$ \ocstrx \xspace for a 6.4Tbps GPU. }

   \label{fig:degin:topo:ubb}
   \vspace{-2ex}
\end{figure}

\begin{figure}[h!t]
    \centering
    \begin{subfigure}[b]{0.22\textwidth}
        \centering
        \includegraphics[height=70pt]{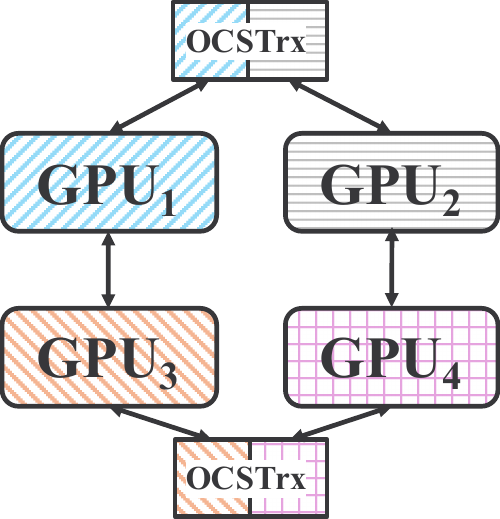}
        \caption{2 bundles of \ocstrx}
        \label{fig:4g3d}
    \end{subfigure}
    \hspace{10pt}
    \begin{subfigure}[b]{0.22\textwidth}
        \centering
        \includegraphics[height=70pt]{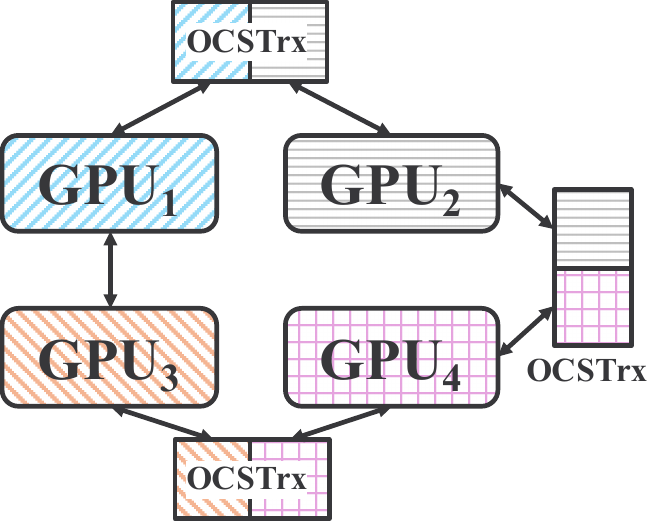}
        \caption{3 bundles of \ocstrx}
        \label{fig:4g2d}
    \end{subfigure}
    \vspace{-2ex}
    \caption{4-GPU node with \docs.}
    \label{fig:inner-topo}
    \vspace{-2ex}

\end{figure}

\para{Intra-node Topology.} The intra-node topology is designed for \textit{dynamic ring construction} and complies with the OCP UBB 2.0 standard~\cite{UBB2.0}. As shown in \figref{fig:degin:topo:ubb}, one node equipped with $R$ GPUs can support $R$ bundles of \ocstrx. Each \docs{} bundle is connected to a pair of GPUs, with one GPU linking to the upper-half SerDes and the other to the lower-half.
For one group of nodes connected as one line, the two GPU \revised{pairs} at each end can interconnect with the \ocstrx \xspace internal loopback path, forming a GPU-level ring. 
As shown in \figref{fig:overview}, nodes $N_1$ and $N_3$ are connected in a line, where $OCSTrx_1(N_1)$ and $OCSTrx_2(N_3)$ activate the cross-lane loopback path, creating a ring between the 8 GPUs of $N_1$ and $N_3$.
During ring construction, only two \docs{} bundles per node are utilized, while the remaining \docs{} operate in loopback mode. These idle \docs{} can be replaced with direct connections, such as DAC links, offering a trade-off between cost and reliability. \figref{fig:inner-topo}(a,b) shows a 4-GPU node with varying numbers of \textit{\docs{}} bundles. Note that the topology design in this section utilizes a 4-GPU node as an example; it can be easily scaled for 8-GPU nodes.

\para{Inter-node Topology.}
We construct the inter-node topology by pruning the full-mesh design, based on two key observations:
i) \textit{Traffic locality}: TP Ring-AllReduce in HBD exhibits neighbor communication patterns, eliminating the need for distant connections; 
ii) \textit{Fault non-locality}: node-side failures typically occur independently at the node level, meaning consecutive multi-node failures follow an exponentially decaying probability.
Each node provides up to $2R$ external paths, allowing us to construct a DC-scale reconfigurable $K$-Hop Ring topology ($K\leq R$) by connecting them to nodes at $\pm 1,...,\pm K$ ($K\leq R$). 
\revised{Under this design, each node has a degree of $2K$, which is sufficient for traffic locality and tunable via OCSTrx count and intra-node design.}
For AllReduce communication, only two out of the $2K$ links are simultaneously activated, with the others serving as backup links for fault isolation.
For example (\figref{fig:overview}), if $N_2$ fails, $OCSTrx_2(N_1)$ and $OCSTrx_1(N_3)$ can switch to backup links, maintaining connectivity between $N_1$ and $N_3$ while isolating $N_2$'s fault. As $K$ increases, the probability of encountering an unbypassed failure rapidly decreases, which is nearly negligible for $K=3$ (detailed analysis in Appendix~\S\ref{appendix:ft-anay}). Thus, this architecture typically achieves a node-level explosion radius.
Moreover, the K-Hop Ring can be broken into the K-Hop line topology, with the trade-off of reduced fault tolerance, affecting the $2K$ nodes at both ends.

\subsection{HBD-DCN Orchestration Algorithm}  
\label{sec:design:orch}  

\sys{} is designed to work with arbitrary DCNs, including Rail-Optimized~\cite{rail-optimized, sigcomm2024hpn} and Fat-Tree~\cite{sigcomm2008fattree}. This section co-optimizes communication performance for both HBD and DCN in \sys{}.

\para{Problem Statement.} In \sys{}, GPUs communicate without routing traffic, preventing congestion at any scale. In contrast, DCNs experience inevitable congestion, leading to performance degradation. To mitigate this, we leverage traffic locality to orchestrate nodes, minimizing cross-ToR traffic. Given a job $J$ requiring $N$ nodes from an available pool of $M$ ($M \geq N$), we must select and order $N$ nodes to satisfy two requirements: (1) nodes in the same TP group should communicate via \sys{}, and (2) other parallel traffic should be arranged to minimize congestion. Ideally, communication remains within the same ToR, confining congestion to switch-to-node links.  

\begin{figure}[h!t]
\centering
\begin{subfigure}[b]{0.23\textwidth}
    \centering
    \includegraphics[width=0.92\textwidth]{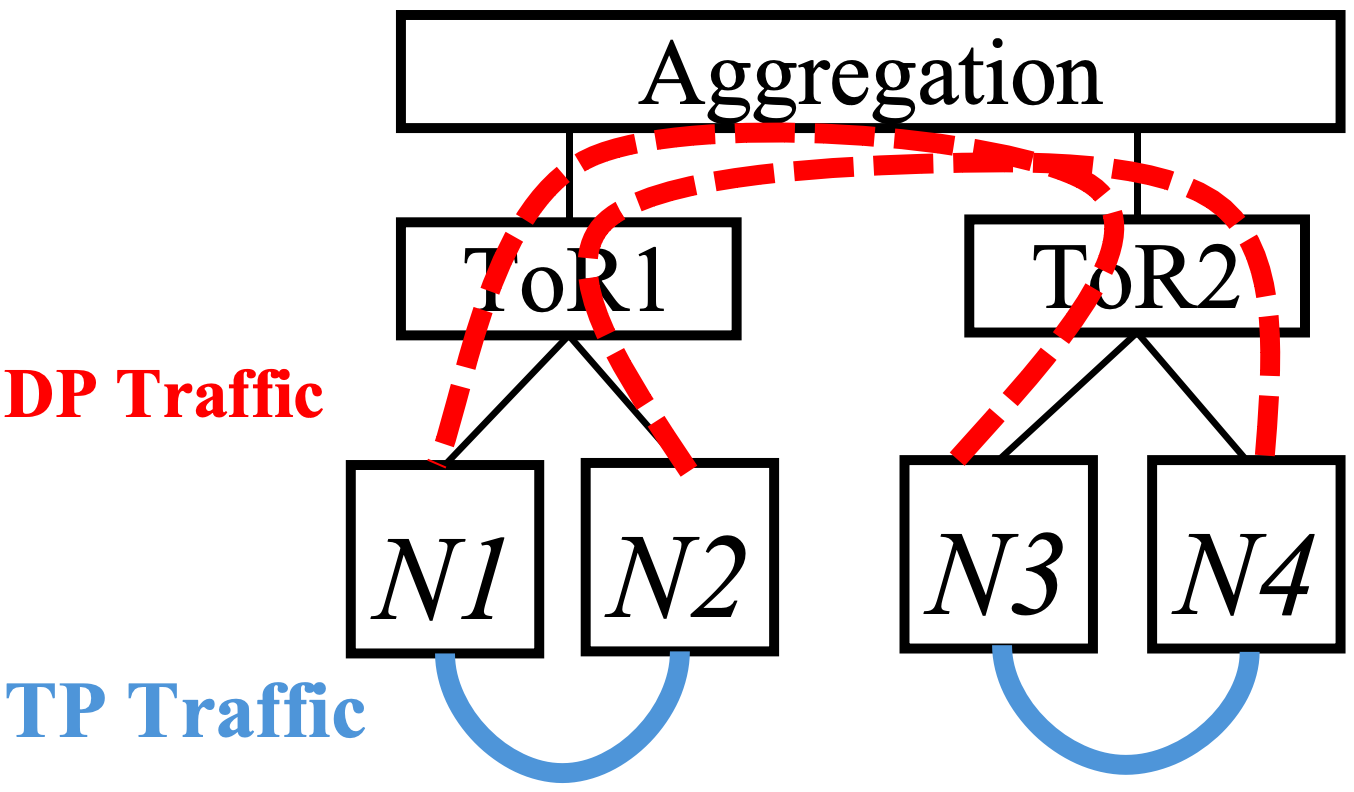}
    \caption{Orchestration scheme 1.}
    \label{figure:orchstration:problem-1}
\end{subfigure}
\hspace{2pt}
\begin{subfigure}[b]{0.23\textwidth}
    \centering
    \includegraphics[width=0.92\textwidth]{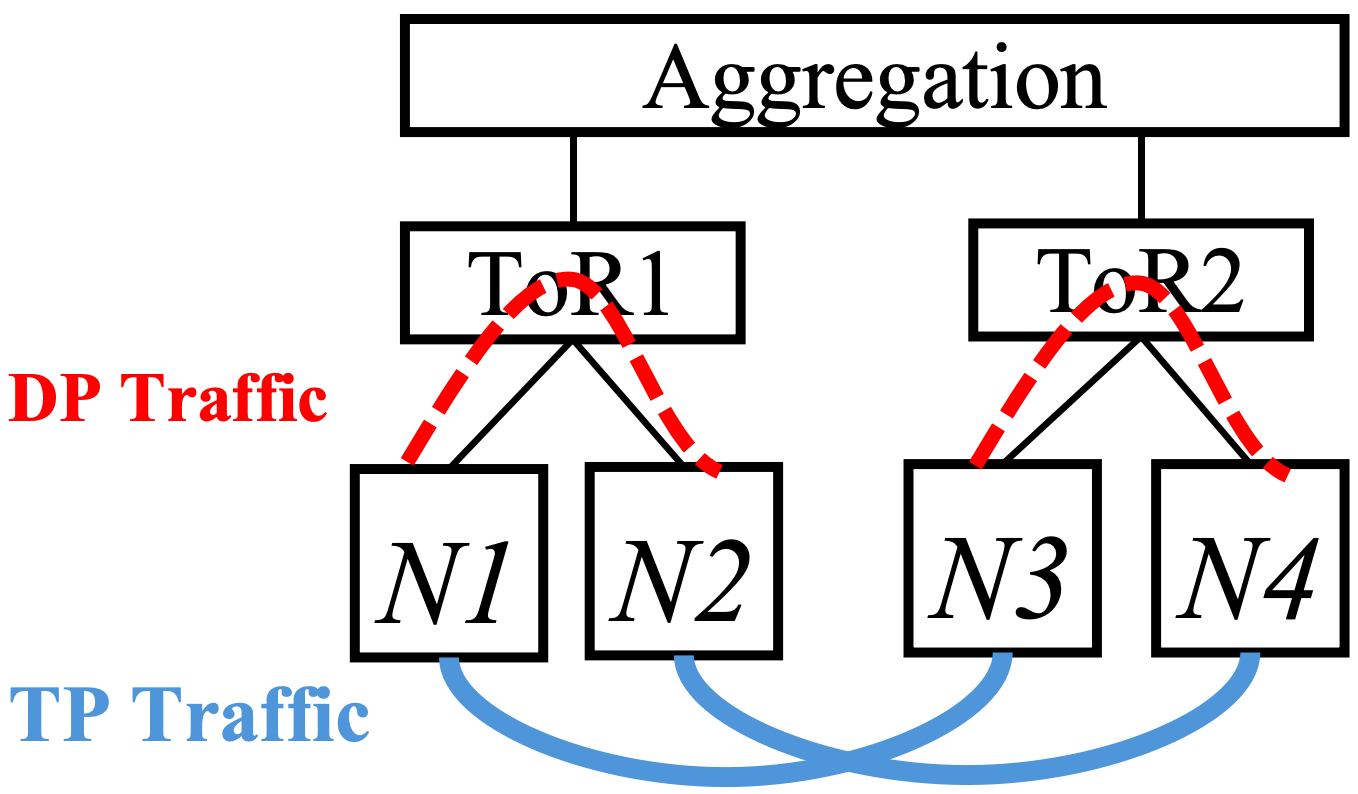}
    \caption{Orchestration scheme 2.}
    \label{figure:orchstration:problem-2}
\end{subfigure}
\caption{Illustration of the node orchestration problem statement.}
\vspace{-1em}
\end{figure}

A naive approach is sorting nodes based on deployment order in \sys{}, fulfilling the first requirement but not the second. As shown in \figref{figure:orchstration:problem-1}, this method places ($N_1$, $N_2$) in the same TP group and ($N_1$, $N_3$) in the same DP group, forcing DP traffic across ToRs. A better scheme (\figref{figure:orchstration:problem-2}) eliminates cross-ToR traffic and congestion. However, considering failures and multiple parallel dimensions complicates orchestration, necessitating an efficient method.

Our key insight is to arrange nodes in \sys{} based on DCN traffic locality, prioritizing appropriate network distances over minimal ones. For example, in \figref{figure:orchstration:problem-2}, $N_1$'s \sys{} neighbor is $N_3$, despite a 3-hop network distance in DCN. We propose a two-phase solution: (1) a deployment phase defining physical connections in DCN and \sys{}, and (2) a runtime phase using an algorithm to orchestrate nodes for arbitrary-scale jobs.

\begin{algorithm}[h!t]
\small
\caption{Orchestration For Fat-Tree}
\label{alg:orchestration-fat-tree-overview}
\SetAlgoNlRelativeSize{-1}
\SetAlgoNlRelativeSize{1}
 \KwIn{
    Topology of DCN and HBD $G$, Faulty Node Set $F$, Job Information $J$.}
 \KwOut{Placement scheme that satisfies job scale and minimizes cross-ToR traffic.}
 Create graph $G_{deploy}=<S_{deploy},E_{deploy}>=\text{Deployment-Strategy}(G)$\;
 Initialize $high=n_{allconstraints}$, $low=0$, $placement =\{\}$\;
\While{ $low \leq high$}
{
     $mid=\lfloor \frac{low+high}{2} \rfloor$\;
     $placement=\text{Placement-Fat-Tree}(G_{deploy},mid,F,J)$\;
    \eIf {$placement$ satisfies job $J$}
    {
         $low=mid+1$\;
    }
    {
         $high=mid-1$\;
    }
}
\KwRet {$placement$}
\end{algorithm}

\begin{figure}[h!t]
    \centering
    \includegraphics[width=0.7\linewidth]{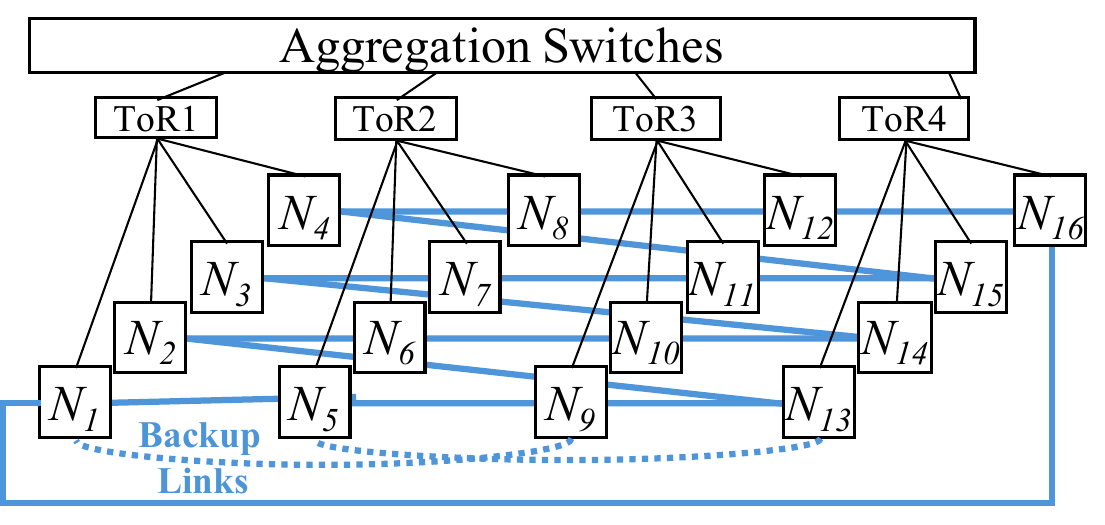}
    \vspace{-1em}
    \caption{Illustration of the deployment phase, showing only two backup links for simplicity.}
    \label{fig:fat-tree-topo}
    \vspace{-8pt}
\end{figure}

\para{Deployment Phase.} \figref{fig:fat-tree-topo} shows node deployment in HBD and DCN. \sys{} connects nodes at a network distance of 3 (i.e., cross-ToR). In a DCN with $r$ nodes per ToR, node $N_n$ connects to $N_{n\pm r}$ as main links, while backup links connect to $N_{n\pm 2r}$. For $1 < n \le r$, $N_n$ connects to $N_{D + n - r - 1}$, where $D$ is the total node count (e.g., $N_3$ connects to $N_{14}$). Additionally, $N_1$ may link to the last node, forming a ring.

\para{Runtime Phase.} Without considering DCN topology, \sys{} orchestrates nodes in three steps: (1) identifying cluster faults and modeling healthy nodes as a graph, (2) using Depth-First Search to find connected components, and (3) sequentially placing TP groups within these components. Due to \sys{}’s topology, each TP group forms a ring.

For real-world DCNs, topology constraints refine step (2) and (3). In Fat-Tree networks, congestion arises when (1) a TP group spans multiple Aggregation-Switch domains, or (2) GPUs within a ToR have mismatched TP group ranks, forcing DP, CP, PP, SP traffic across ToRs. Thus, we aim to localize TP groups within the same Aggregation-Switch domain and align ranks within each ToR. Our scheduling algorithm minimizes cross-ToR traffic while meeting job scale requirements via a binary search over constraint variables. \algref{alg:orchestration-fat-tree-overview} outlines the approach, with full details in Appendix~\S\ref{appendix:orch-algo}.

\section{Hardware and Small-Scale Cluster Evaluation}
\label{sec:hardware-system}

\subsection{\docs{} Hardware Evaluation}
\label{sec:hardware-system:hardware}

\begin{figure}[h!t]
    \centering
    \begin{subfigure}[b]{0.23\textwidth}
        \centering
        \includegraphics[height=40pt]{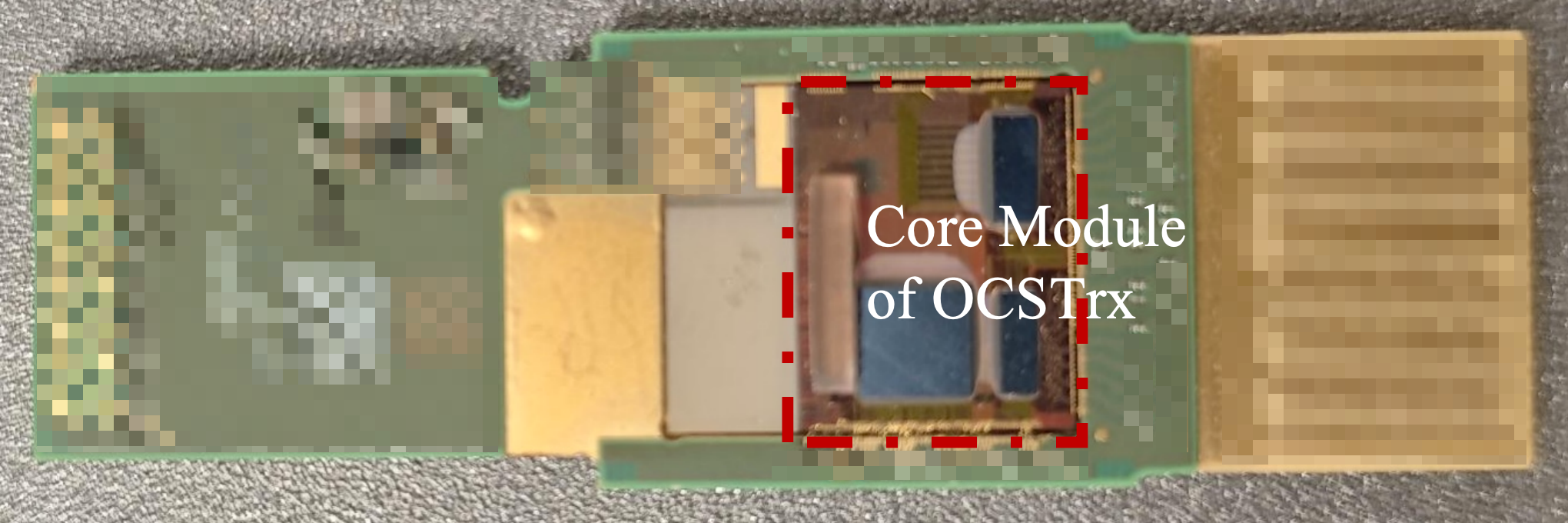}
        \caption{\revised{Main hardware components.}}
        \label{fig:hardware:ocstrx-impl}
    \end{subfigure}
    \hspace{2pt}
    \begin{subfigure}[b]{0.23\textwidth}
        \centering
        \includegraphics[height=40pt]{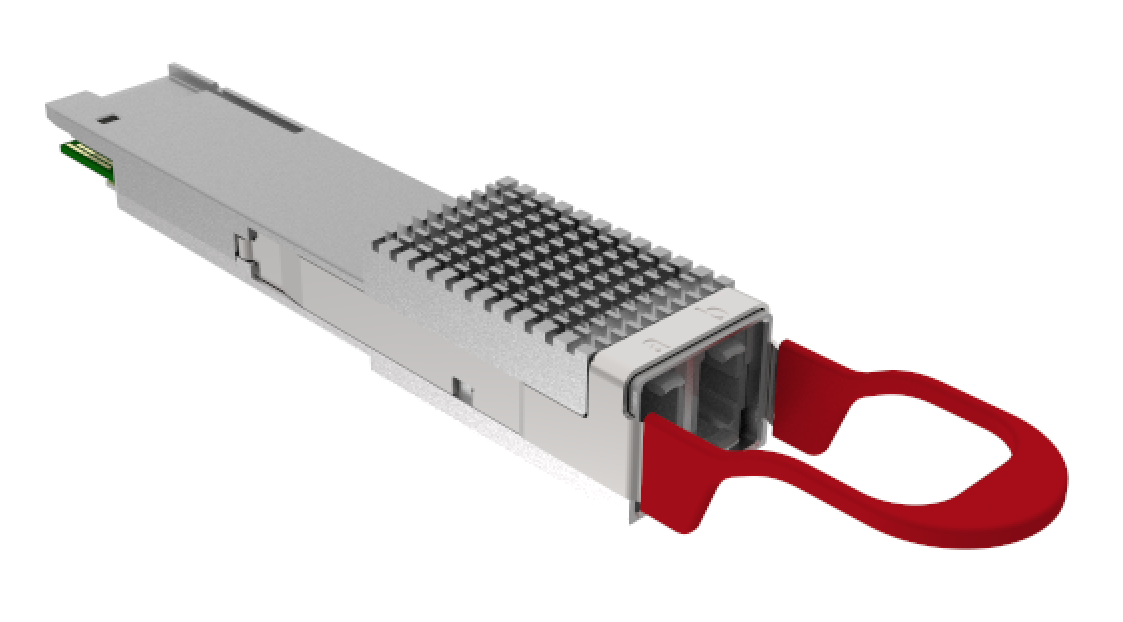}
        \caption{\revised{Packaged module.}}
        \label{fig:hardware:ocstrx-embedded}
    \end{subfigure}
    \vspace{-1em}
    \caption{\revised{\ocstrx{} integrated in QSFP-DD 800Gbps transceiver.}}
    \label{fig:hardware}
\end{figure}

\begin{figure}[h!t]
    \vspace{-1em}
    \centering
    \includegraphics[width=0.48\textwidth]{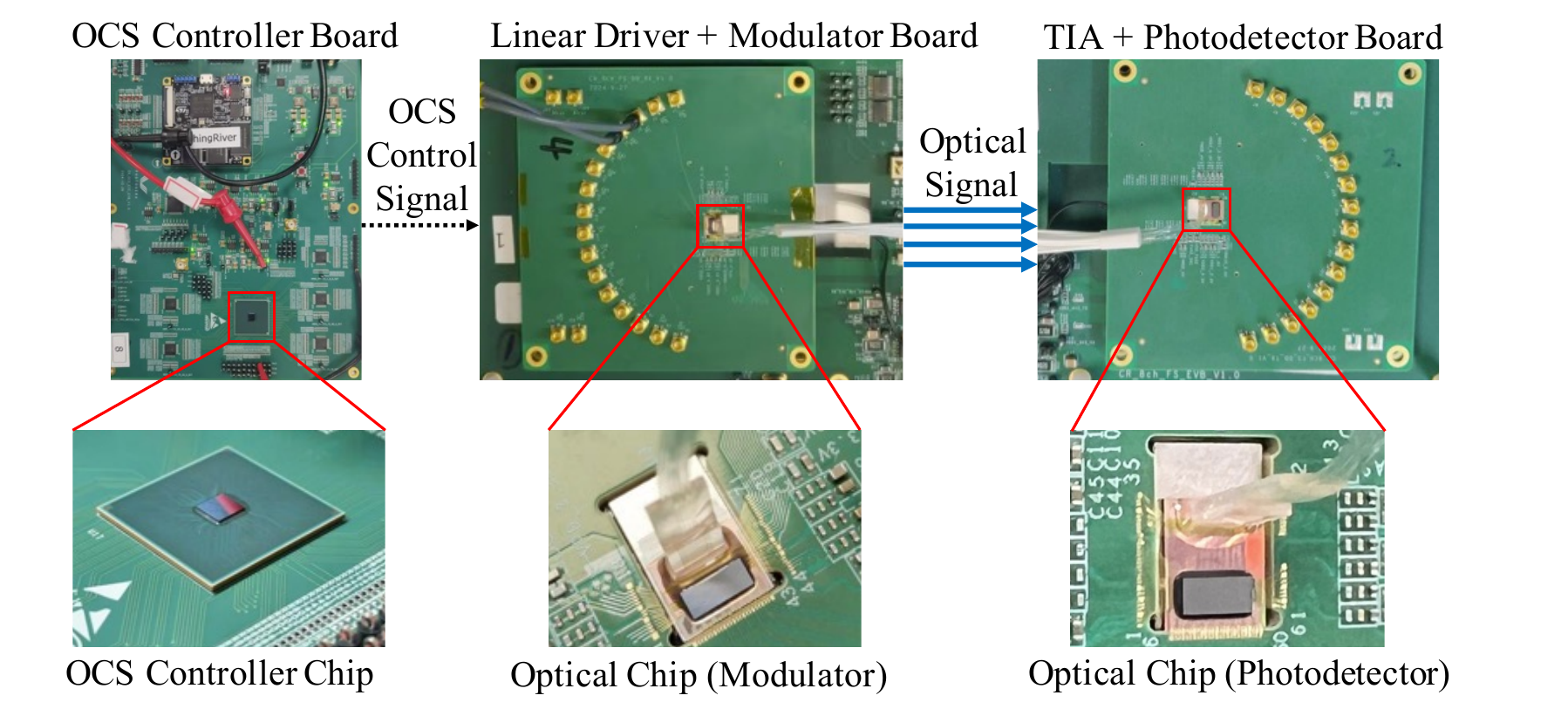}
    \vspace{-20pt}
    \caption{Evaluation board for components of \docs.}
    \label{figure:design:evaluation-board}
    \vspace{-1em}
\end{figure}

\revised{
Following the design principles outlined in \S\ref{sec:design:docs}, we have successfully implemented \ocstrx{}, a fully integrated module within a QSFP-DD 800Gbps transceiver, as illustrated in \figref{fig:hardware}.}
As depicted in \figref{figure:design:evaluation-board}, \ocstrx{} integrates an OCS Controller Chip and a Photonic Integrated Circuit (PIC) that includes an MZI switch matrix. The Controller Chip, measuring $4mm \times 4mm$, is manufactured using a 28nm process, while the PIC, sized at $10.5mm \times 13mm$, uses a 65nm CMOS process. \ocstrx{} supports 8 pairs of TX/RX SerDes at each end and has been validated for compatibility with various link layer protocols, including PCIe (32Gbps, 64Gbps) and Ethernet (56Gbps, 112Gbps).
\revised{\ocstrx{} has also passed temperature cycling tests, validating its reliability and availability.}

\para{\revised{Reconfiguration latency.}}
\revised{Experimental results show that \ocstrx{} achieves a reconfiguration latency of 60–80 $\mu s$, which is significantly faster than traditional centralized OCS-based switches employing MEMS~\cite{missionapollo}, Piezo~\cite{piezo-ocs}, or Robotic~\cite{robotics-ocs} mechanisms, whose latencies typically range from milliseconds to minutes. Notably, this measurement excludes software-level delays such as reconnection at the network protocol layer.}

\begin{figure}[h!t]
    \centering
    \begin{subfigure}[b]{0.23\textwidth}
        \centering
        \includegraphics[width=\textwidth]{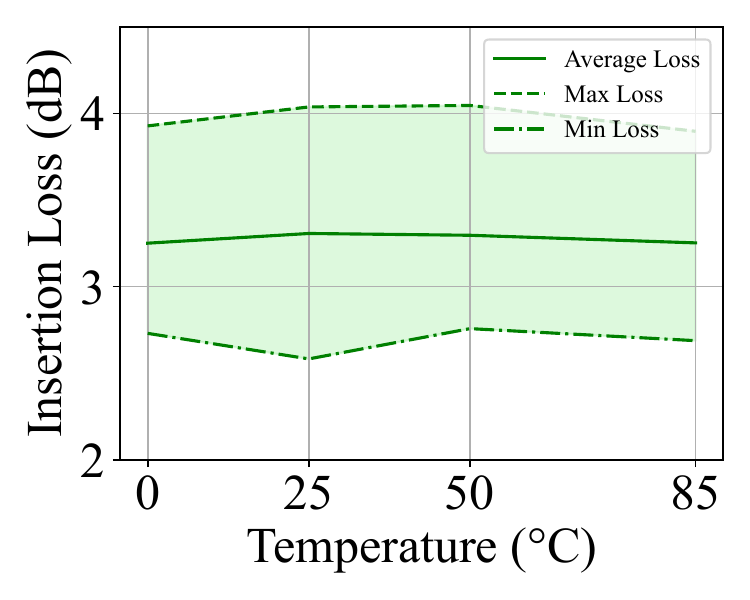}
        \caption{\revised{Insertion loss.}}
        \label{fig:hardware:loss-power:loss}
    \end{subfigure}
    \hspace{2pt}
    \begin{subfigure}[b]{0.23\textwidth}
        \centering
        \includegraphics[width=\textwidth]{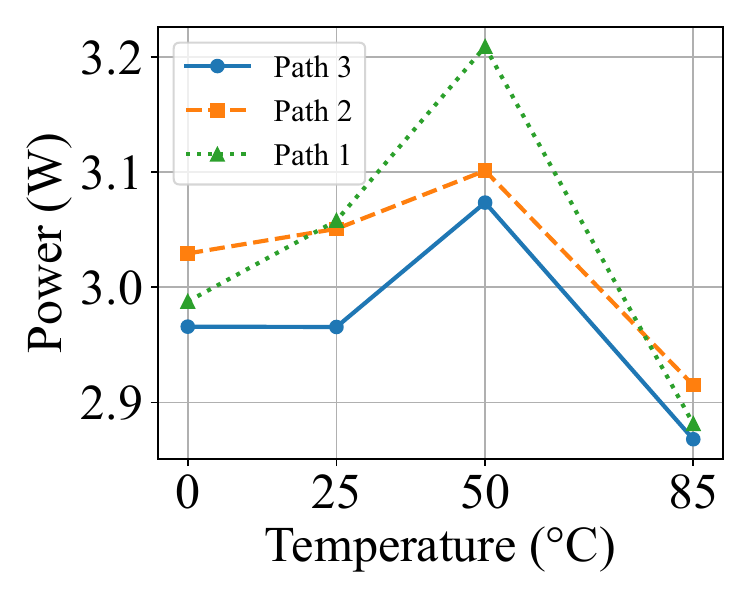}
        \caption{\revised{Power consumption.}}
        \label{fig:hardware:loss-power:power}
    \end{subfigure}
    \vspace{-10pt}
    \caption{\revised{Insertion loss and power consumption of the core module in \ocstrx{}.}}
    \label{fig:hardware:loss-power}
\end{figure}

\begin{figure*}[h!t]
    \centering
    \begin{subfigure}[b]{0.23\linewidth}
        \centering
        \includegraphics[width=\linewidth]{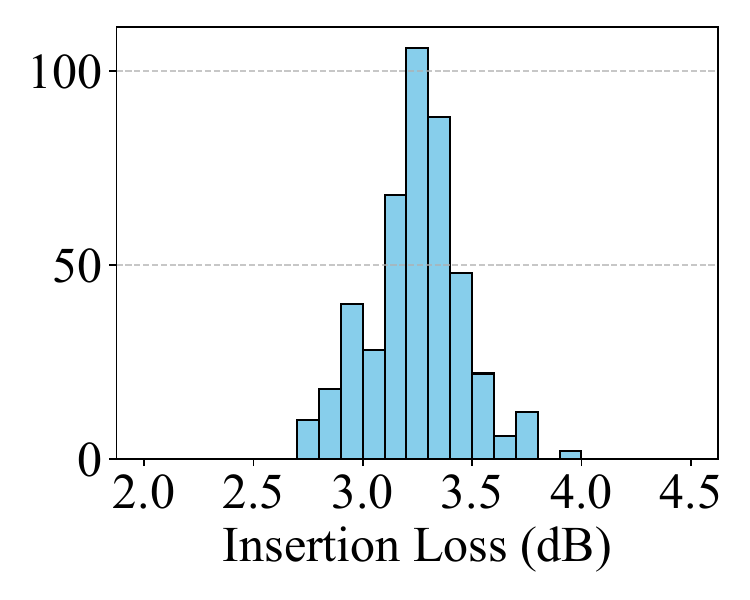}
        \caption{\revised{0°C.}}
        \label{fig:hardware:loss:distribution:0}
    \end{subfigure}
    \hspace{2pt}
    \begin{subfigure}[b]{0.23\linewidth}
        \centering
        \includegraphics[width=\linewidth]{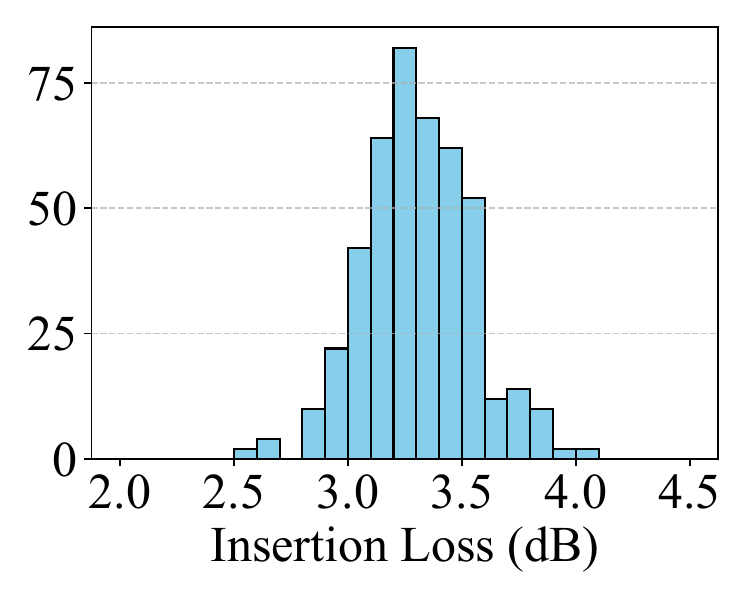}
        \caption{\revised{25°C.}}
        \label{fig:hardware:loss:distribution:25}
    \end{subfigure}
    \hspace{2pt}
    \begin{subfigure}[b]{0.23\linewidth}
        \centering
        \includegraphics[width=\linewidth]{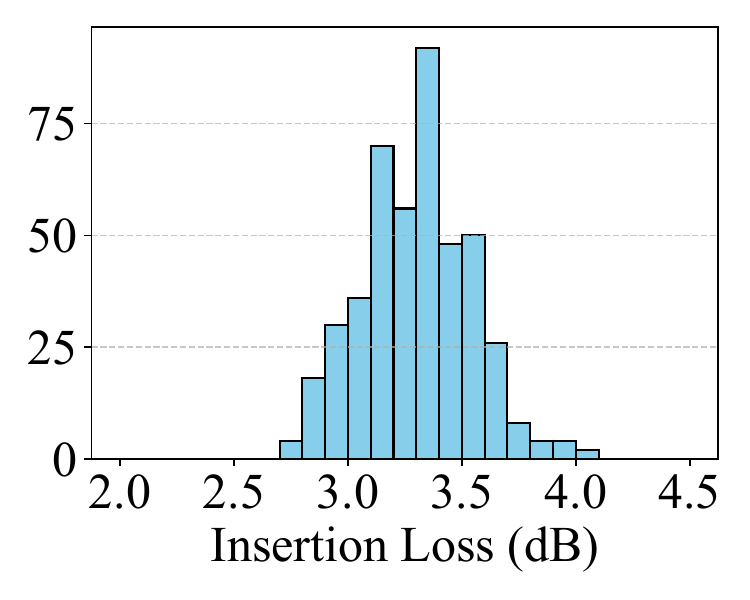}
        \caption{\revised{50°C.}}
        \label{fig:hardware:loss:distribution:50}
    \end{subfigure}
    \hspace{2pt}
    \begin{subfigure}[b]{0.23\linewidth}
        \centering
        \includegraphics[width=\linewidth]{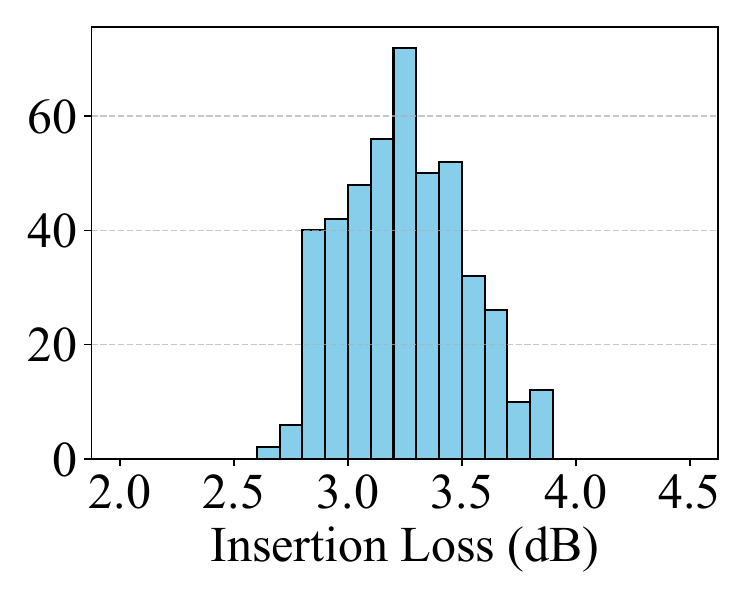}
        \caption{\revised{85°C.}}
        \label{fig:hardware:loss:distribution:85}
    \end{subfigure}
    \vspace{-1ex}
    \caption{\revised{Insertion loss distribution of the core module in \ocstrx{} under different ambient temperatures.}}
    \label{fig:hardware:loss:distribution}
\end{figure*}

\para{\revised{Insertion loss.}}
\revised{We evaluated the insertion loss of the core module within \ocstrx{}, which provides the OCS capability, under various ambient temperatures. \figref{fig:hardware:loss-power:loss} summarizes the statistical results, while \figref{fig:hardware:loss:distribution} shows the distribution. The measured insertion loss ranges from 2.5 dB to 4.0 dB, with an average of 3.3 dB at room temperature (25°C).}

\para{\revised{Power consumption.}}
\revised{Under the $8 \times 112G$ configuration, the peripheral circuitry consumes 8.5 $Watts$. We further measured the power consumption of \ocstrx{}’s core module which provides the OCS capability across different temperatures when activating three paths (two external paths and one internal loopback), as illustrated in \figref{fig:hardware:loss-power:power}. Across all conditions, the core module consumed less than 3.2 $Watts$, keeping the total power below 12 $Watts$, meeting the QSFP-DD 800Gbps specification \cite{qsfp-dd-15w}.}

\begin{figure*}[h!t]
    \centering
    \begin{subfigure}[b]{0.23\linewidth}
        \centering
        \includegraphics[width=\linewidth]{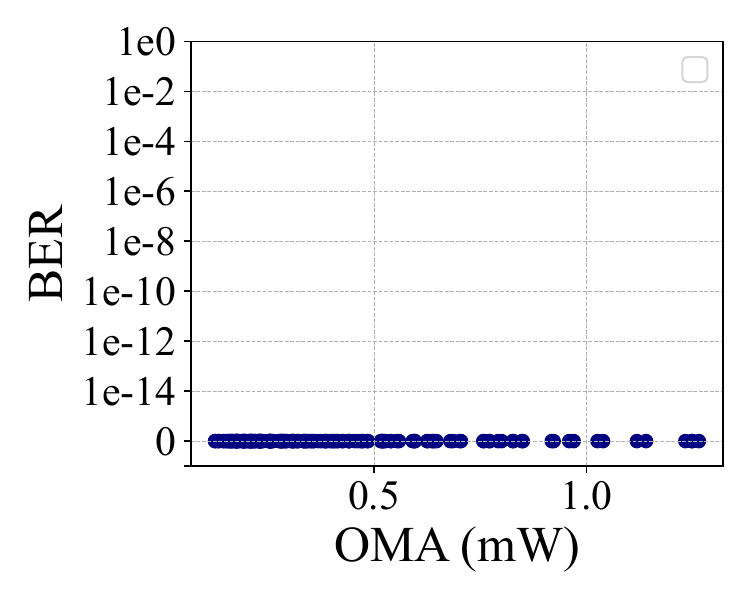}
        \caption{\revised{-5°C.}}
        \label{fig:hardware:ber:-5}
    \end{subfigure}
    \hspace{2pt}
    \begin{subfigure}[b]{0.23\linewidth}
        \centering
        \includegraphics[width=\linewidth]{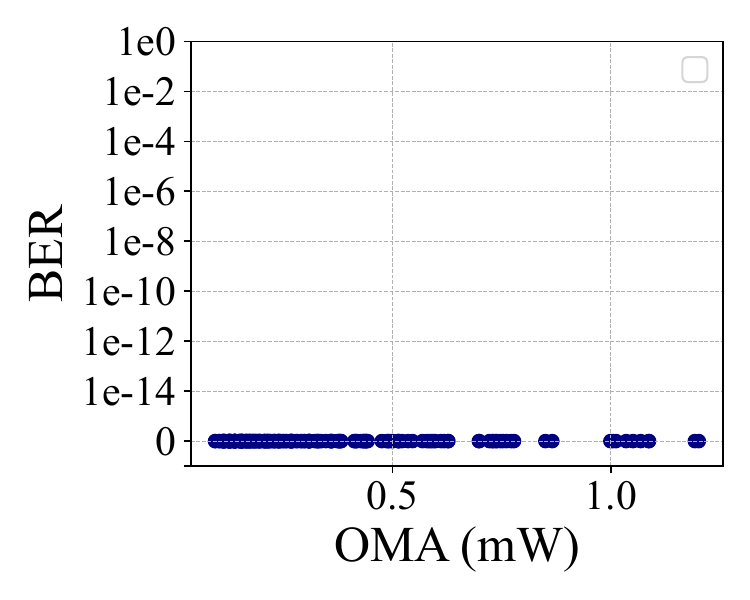}
        \caption{\revised{25°C.}}
        \label{fig:hardware:ber:25}
    \end{subfigure}
    \hspace{2pt}
    \begin{subfigure}[b]{0.23\linewidth}
        \centering
        \includegraphics[width=\linewidth]{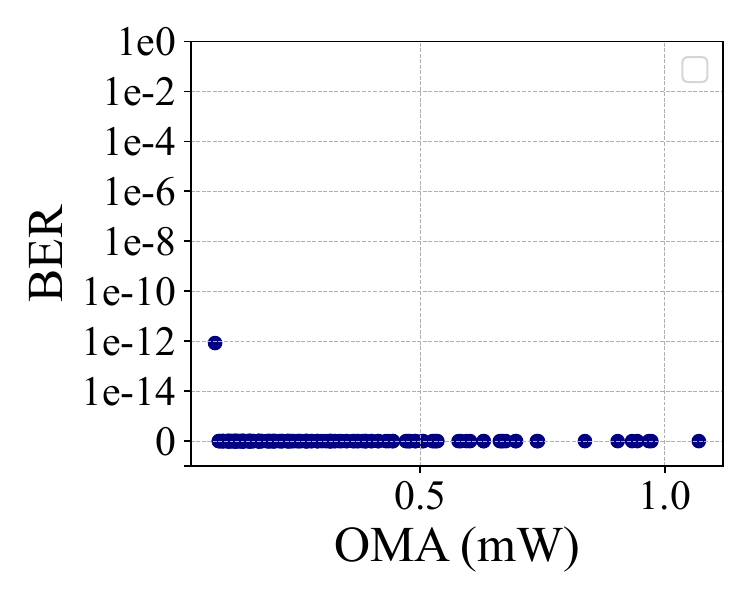}
        \caption{\revised{50°C.}}
        \label{fig:hardware:ber:50}
    \end{subfigure}
    \hspace{2pt}
    \begin{subfigure}[b]{0.23\linewidth}
        \centering
        \includegraphics[width=\linewidth]{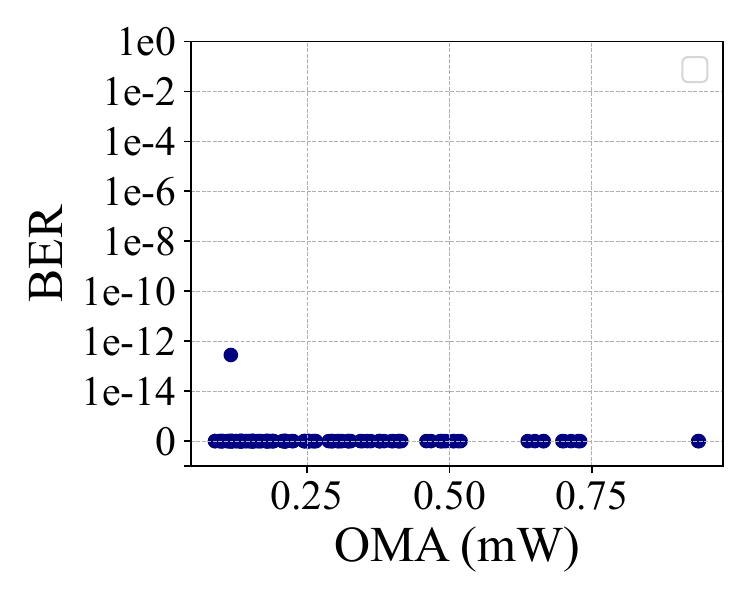}
        \caption{\revised{75°C.}}
        \label{fig:hardware:ber:75}
    \end{subfigure}
    \vspace{-1ex}
    \caption{\revised{Bit error rate of \ocstrx{} under varying OMA and ambient temperatures.}}
    \label{fig:hardware:ber}
\end{figure*}

\para{\revised{Bit error rate.}}
\revised{With the Amplitude Control (AC) driving power set to 620 mW, we measured the bit error rate (BER) of \ocstrx{} under different optical modulation amplitudes (OMAs) and ambient temperatures, as shown in \figref{fig:hardware:ber}. At -5°C and 25°C, BER was consistently 0. At 50°C and 75°C, BER remained 0 in most cases, with occasional errors only at very low OMAs. These results confirm compliance with the industrial BER threshold.}

\subsection{Small-Scale Cluster Evaluation}
\label{sec:testbed:minipod}
We constructed a small-scale cluster to evaluate the communication performance of the ring topology. Using 32 experimental GPUs equipped with inter-host HBD support (96 lanes on PCIe 4 protocol), we formed a physical ring utilizing fixed optical modules. This mini-cluster was manually reconfigured for both 32-GPU and 16-GPU ring topologies. The communication latency and AllReduce performance are evaluated.
For small packets, direct GPU-to-GPU links reduced latency by approximately 13\% compared to the NVLink switch design.
For large packets, the 16-GPU AllReduce utilized 77.11\% of the ring bandwidth, with the utilization rate increasing to 77.26\% for the 32-GPU configuration, showing minimal degradation with scaling. In comparison, the NVIDIA H100 8-GPU machine achieves an 81.77\% utilization rate without SHARP.

\revised{Additionally, a control plane is included to manage the \sys{}. At the device level, the \textbf{node fabric manager} configures individual \ocstrx{} modules and handles topology switching. At the system level, \textbf{cluster manager} coordinates global control across the cluster.
As control-plane mechanisms are not the focus of this work, we omit further details.}

\section{Large-Scale Simulation}
\label{sec:simulation}

We begin by outlining the experimental methodology and setup (\S\ref{sec:simulation:setup}). Next, we assess fault tolerance across different HBD architectures (\S\ref{sec:simulation:fault}), followed by end-to-end simulations to evaluate training performance under varying parallelism and GPU resource allocations (\S\ref{sec:simulation:end2end}). We then examine the improvements in communication efficiency achieved by our orchestration algorithm (\S\ref{sec:simulation:efficiency}). Finally, we present a comparative cost and power analysis of different HBD architectures (\S\ref{sec:simulation:cost-power}). The simulations demonstrate that \sys{} outperforms other architectures across all metrics.

\subsection{Methodology and Setup}
\label{sec:simulation:setup}

An in-house simulator dedicated to LLM training is used to evaluate \sys comprehensively. The simulator supports end-to-end simulations of both model training performance and hardware faults, with the HBD-DCN orchestration algorithm seamlessly integrated into the system.

\begin{figure}[h!t]
    \centering
    \begin{subfigure}[b]{0.23\textwidth}
        \centering
        \includegraphics[width=\textwidth]{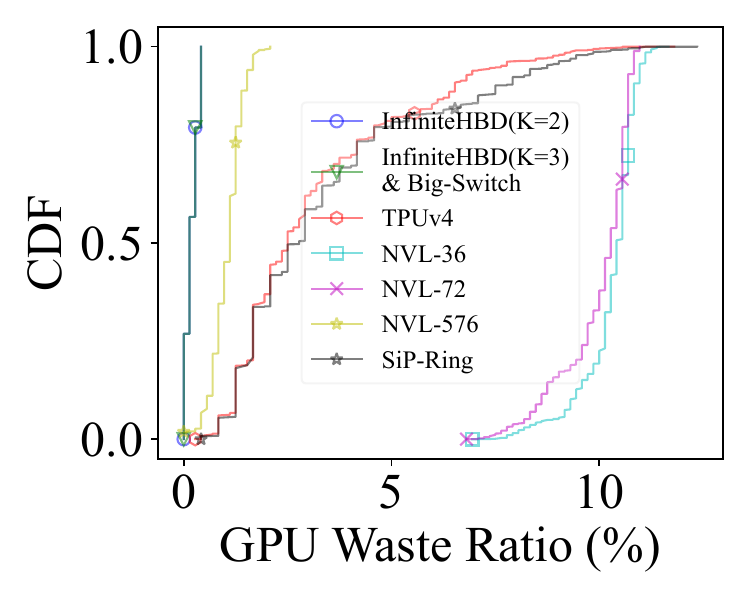}
        \caption{TP-16.}
        \label{fig:simulation:waste-cdf:tp16-gr8}
    \end{subfigure}
    \hspace{2pt}
    \begin{subfigure}[b]{0.23\textwidth}
        \centering
        \includegraphics[width=\textwidth]{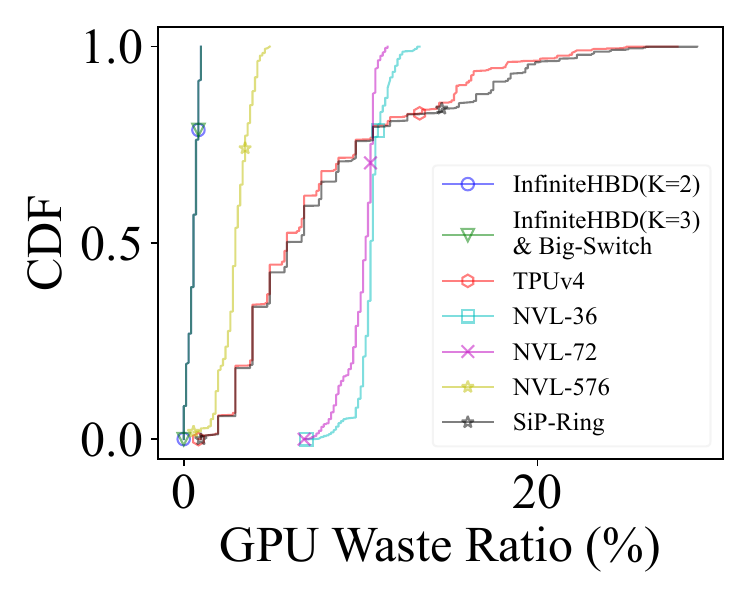}
        \caption{TP-32.}
        \label{fig:simulation:waste-cdf:tp32-gr8}
    \end{subfigure}
    \caption{CDF of GPU waste ratio over 4-GPU node based on production fault trace.}
    \label{fig:simulation:waste-cdf:gr4}
\end{figure}

\para{GPU and network specification.}
The NVIDIA H100~\cite{h100} (989 TFLOPS, 80GiB) is used as the configuration of GPU in simulation. The HBD bandwidth of GPU is set to $6.4Tbps$, which is the sum of 8 QSFP-DD \ocstrx. The DCN bandwidth is configured to match the NVIDIA ConnectX-7 ($400Gbps$). 
Since the simulation primarily focuses on HBD, the DCN is configured as a Fat-Tree topology~\cite{sigcomm2008fattree}. Several HBD architectures are then evaluated, including:

\begin{itemize}[itemsep=2pt,topsep=0pt,parsep=0pt, leftmargin=2ex]
    \item \textbf{Big-Switch}: The ideal HBD design, featuring a large centralized switch with no forwarding latency that connects all nodes, as the theoretical upper limit of communication performance and fault resilience.
    \item \textbf{\sys{}}: Two configurations are evaluated: the \ocstrx{} bundle is set to either $K = 2$ or $K = 3$ (\S\ref{section:design:topology}), constructing 2/3-Hop Ring respectively.
    \item \textbf{NVL-36, NVL-72, NVL-576}~\cite{nvl72}: HBDs with 36, 72, or 576 GPUs, GPUs are interconnected via NVLink Switches.
    \item \textbf{TPUv4}~\cite{isca2023tpu}: Centralized OCS capable of scheduling with a $4^3$ TPU cube granularity.
    \item \textbf{SiP-Ring}~\cite{sip-ml}: All nodes are connected in a series of static rings with fixed sizes equal to the TP sizes.
\end{itemize}

\para{GPU count per node. }The simulation aligns with both 4-GPU node (e.g., NVIDIA GB200 NVL-36/72/576~\cite{nvl72} and TPUv4~\cite{isca2023tpu}) and 8-GPU node designs (NVIDIA H100, AMD MI300X~\cite{amdmi300}, Intel Gaudi3~\cite{intelgaudi3}, and UBB 2.0 standard~\cite{UBB2.0}).

\begin{figure}[h!t]
    \centering
    \begin{subfigure}[b]{0.23\textwidth}
        \centering
        \includegraphics[width=\textwidth]{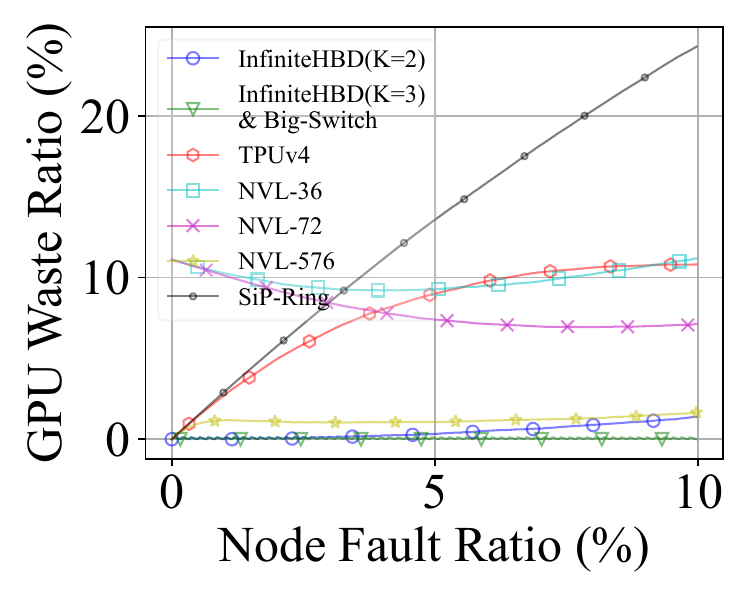}
        \caption{TP-16.}
        \label{fig:simulation:model:wasted-overview:tp16}
    \end{subfigure}
    \hspace{2pt}
    \begin{subfigure}[b]{0.23\textwidth}
        \centering
        \includegraphics[width=\textwidth]{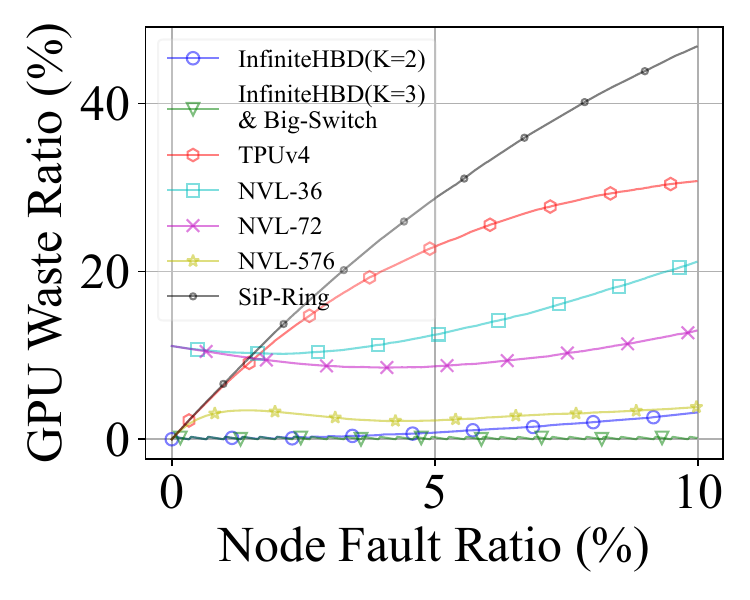}
        \caption{TP-32.}
        \label{fig:simulation:model:wasted-overview:tp32}
    \end{subfigure}
    
    \caption{GPU wastes ratio over the 4-GPU node with different GPU fault ratio based on fault model.}
    \label{fig:simulation:model:wasted-overview}
\end{figure}

\para{Parallelism strategy. } 
Since \sys is primarily designed for TP, the key variable is the TP size. TP-8, TP-16, TP-32, and TP-64 are tested to evaluate the fault resilience of various HBD architectures (\S\ref{sec:simulation:fault}).
Additionally, other parallelism strategies, such as PP and DP, are used to simulate cross-ToR traffic and evaluate the orchestration algorithm (\S\ref{sec:simulation:efficiency}).

\para{Fault patterns. } The fault trace used in the simulation was collected from an 8-GPU node cluster with approximately 3K-GPUs over 348 days.
On average, the ratio of faulty 8-GPU nodes is $2.33\%$, with the P99 value as $7.22\%$, more details in Appendix~\S\ref{appendix:production-fault-trace}. In some simulations, fault traces generated based on this trace statistics are also derived. 

\subsection{HBD Fault Resilience}
\label{sec:simulation:fault}

This section evaluates the fault resilience of various HBD architectures, focusing on GPU waste ratio, job fault-waiting time, and the maximum job scale supported by the cluster. The main text presents the key results, with more detailed results provided in Appendix~\S\ref{appendix:wasted-GPUs-ratio}.

\para{GPU waste.} 
Apart from faulty GPUs, issues such as fragmentation, topology disconnections, and bandwidth degradation can render healthy GPUs wasted.
The GPU waste ratio quantifies the number of wasted GPUs under different fault scenarios. \figref{fig:simulation:waste-cdf:gr4} illustrates GPU waste ratios over production trace, while \figref{fig:simulation:model:wasted-overview} depicts the GPU waste ratio as node fault ratio varies.

\begin{table}[h!t] \small
    \centering
    \begin{tabular}{llllll}
    \toprule
    \textbf{GPU Num} & \textbf{TP} & \textbf{DP} & \textbf{PP} & \textbf{EP} & \textbf{MFU} \\
    \midrule
    1024    & 16       & 16      & 4       & 1       & 0.4276         \\
    2048    & 16      & 16      & 8        & 1       & 0.4140        \\
    4096    & 32      & 16      & 8        & 1       & 0.3894        \\
    8192    & 32      & 16      & 16      & 1       & 0.3656       \\
    16384  &  64     & 16       & 16      & 1      & 0.3116       \\
    \bottomrule
    \end{tabular}
    \caption{Optimal parallelism strategies for maximizing MFU of GPT-MoE under varying GPU numbers.}
    \vspace{-2em}
    \label{tab:eval:gpt-moe-optimal}
\end{table}

In these scenarios, \sys{} ($K=3$) achieves near-zero GPU waste ratio, and outperforms all other architectures. Especially, the waste ratio for \sys ($K=2$) remains almost identical to that of \sys{} ($K=3$), allowing one bundle of \docs{} to be saved for clusters with low fault rates.   
NVL-36 and NVL-72 typically experience an 11\% waste ratio for TP sizes of 16 or larger, as $1/9$ of GPUs are reserved for redundant backups. NVL-576 has less fragmentation, benefiting from its larger size. TPUv4 performs well at low fault ratios and small TP sizes, but significantly degrades with larger TP sizes due to its coarse $4^3$ cube-based resource management, which amplifies the fault explosion radius. To sum up, \sys{} demonstrates the strongest fault resilience among all architectures.

\begin{figure}[h!t]
    \centering
        \includegraphics[width=0.8\linewidth]{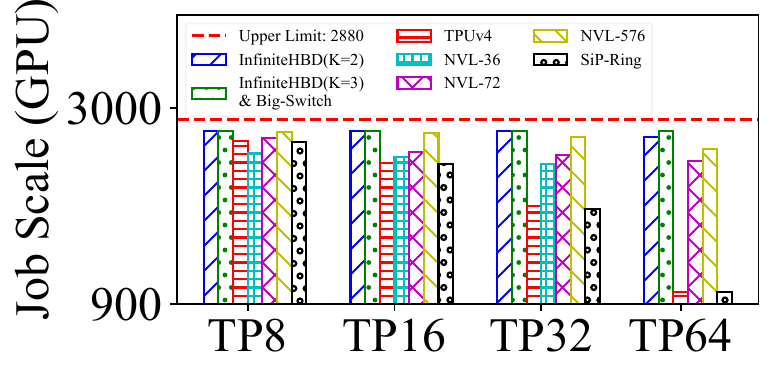}
    \caption{Maximal job scale supported by 2,880 GPUs.}
    \label{fig:simulation:job_scale}
    \vspace{-1em}
\end{figure}

\para{Maximum job supported. } 
In fixed-size clusters, large jobs must pause when the available GPUs drop below the required count. Faced with the same fault rate, clusters with lower GPU waste ratio can support larger job scales. \figref{fig:simulation:job_scale} shows the maximum job scale supported for various HBD architectures cluster with 2,880-GPU, simulated with the fault traces normalized for 4-GPU nodes. \sys{} ($K=2$ or $K=3$) and NVL-576 lead in performance, and SiP-Ring exhibits declining efficiency as TP size increases.

\begin{figure}[h!t]
    
    \centering
    \begin{subfigure}[b]{0.23\textwidth}
        \centering
        \includegraphics[width=\textwidth]{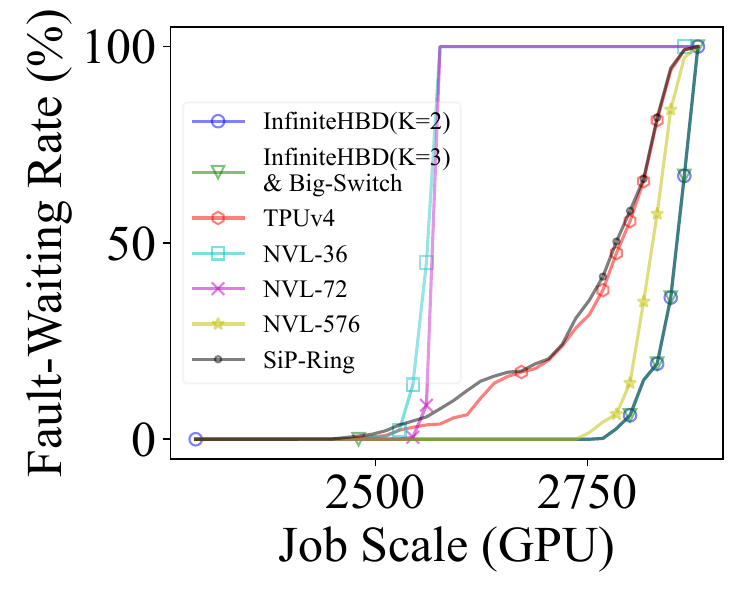}
        \vspace{-1em}
        \caption{TP-16.}
        \label{fig:simulation:breakdown-duration:tp16-8gpu}
    \end{subfigure}
    \hspace{2pt}
    \begin{subfigure}[b]{0.23\textwidth}
        \centering
        \includegraphics[width=\textwidth]{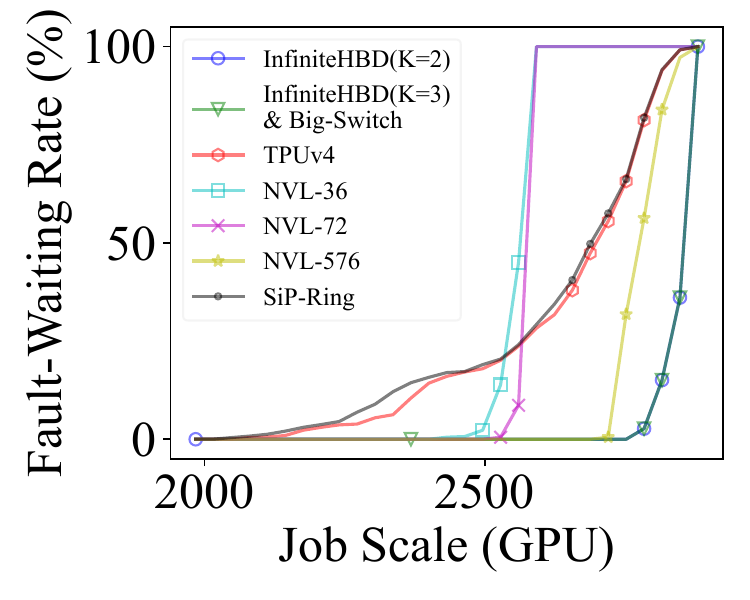}
        \vspace{-1em}
        \caption{TP-32.}
        \label{fig:simulation:breakdown-duration:tp32-8gpu}
    \end{subfigure}
    \vspace{-1em}
    \caption{Job fault-waiting time over the 4-GPU node with different levels of job-scale.}
    \vspace{-1em}
    \label{fig:simulation:breakdown-duration}
\end{figure}

\para{Job fault-waiting time.} Large jobs must wait for repairs when GPU availability falls below the required threshold. These simulations assume the average recovery time in the fault trace as a fixed repair duration. The total wasted time during 348 days is evaluated (\figref{fig:simulation:breakdown-duration}). For smaller TP sizes (TP-8/TP-16), NVL-36/NVL-72 exhibit the weakest resilience due to their 11\% backup overhead. For larger TP sizes (TP-32/TP-64), SiP-Ring and TPUv4 perform worst.

\subsection{Training Performance}
\label{sec:simulation:end2end}

This section analyzes the training performance of two representative large models, {LLama 3.1-405B}~\cite{llama3herdmodels} and {GPT-MoE} (configuration detailed in Appendix~\S\ref{appendix:gpt-moe}), under various GPU resource configurations and parallelism strategies. The simulation results validate the practical applicability of the \sys{} architecture. In simulations, we model practical TP and EP behaviors: For TP, increasing parallelism splits GEMMs into smaller, less efficient tasks, reducing hardware efficiency~\cite{gemm-eff}; for EP, we practically set expert imbalance coefficient at 20\%.

\para{LLama 3.1-405B\footnote{To support larger-scale TP parallelism, we simplified the GQA~\cite{GQA} architecture of LLama 3.1-405B to a traditional MHA architecture.}. }The model adopts a classic decoder-only Transformer architecture. The simulation employs the conventional 3D parallelism strategy\footnote{$TP \in \{1,2,4,8,...,128\}$, $DP \in \{1,2,4,8,...,1024\}$, $PP \in \{1,2,4,8,16\}$, $bsz=2048$}, which combines TP, DP, and PP for performance analyses. 
\tabref{tab:eval:llama3-optimal} presents the optimal parallelism strategies and their corresponding MFU for LLama 3.1-405B under varying GPU resources. As GPU resources increase, the optimal TP size also increases. When the number of GPUs exceeds 8192, the traditional 8-GPU HBD architecture within a single node begins to limit training efficiency. As the cluster size expands, larger TP sizes become increasingly optimal.

\para{GPT-MoE.} The model utilizes the Mixture-of-Experts (MoE) architecture, with $EP \in \{1,2,4,8\}$ introduced in the simulation. \tabref{tab:eval:gpt-moe-optimal} shows the optimal parallelism strategy and the corresponding MFU for GPT-MoE under various GPU resources. The optimal EP value is 1, suggesting that MoE can also achieve high efficiency with TP.

\vspace{-1ex}
\subsection{Communication Efficiency}
\label{sec:simulation:efficiency}

\begin{figure*}[!t]
    \centering
    \hfill{}
    \begin{subfigure}[b]{0.23\textwidth}
        \centering
        \includegraphics[width=\textwidth]{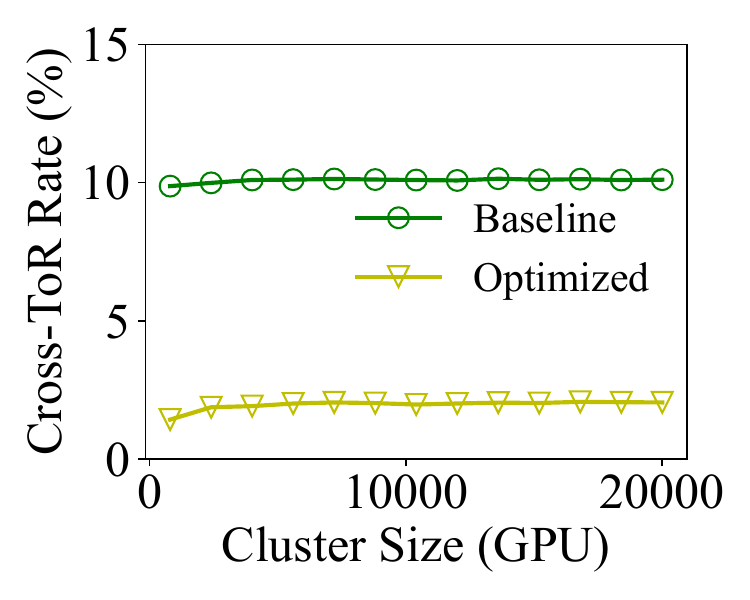}
        \caption{Sensitivity to cluster size.}
        \label{fig:simulation:orch:cluster}
    \end{subfigure}
    \hfill{}
    \begin{subfigure}[b]{0.23\textwidth}
        \centering
        \includegraphics[width=\textwidth]{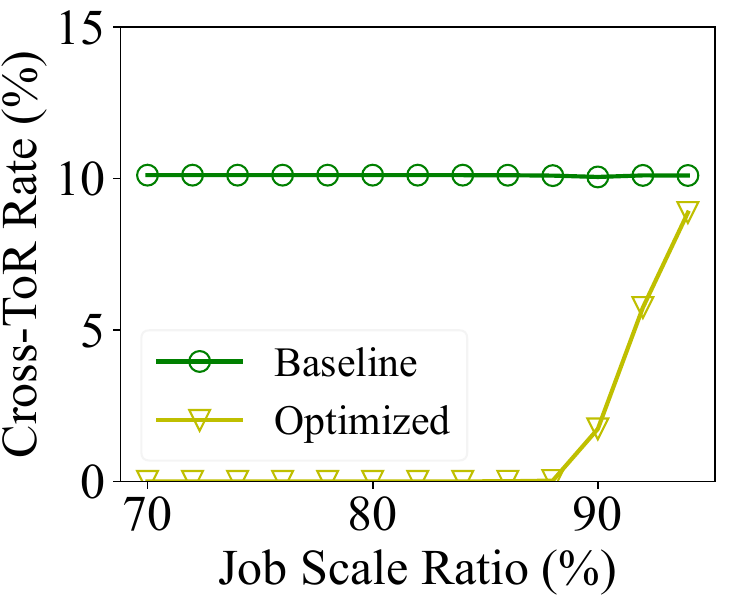}
        \caption{Impact of job-scale ratio.}
        \label{fig:simulation:orch:job}
    \end{subfigure}
    \hfill{}
    \begin{subfigure}[b]{0.23\textwidth}
        \centering
        \includegraphics[width=\linewidth]{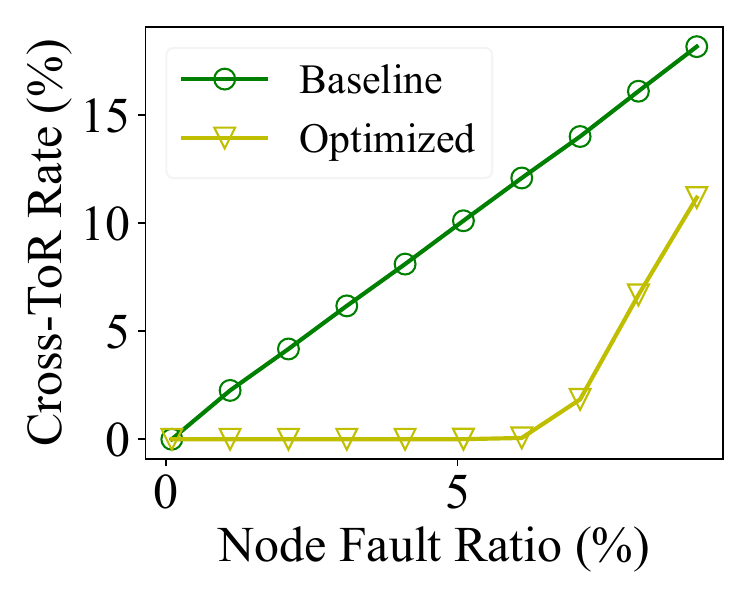}
        \caption{Sensitivity to fault ratio.}
        \label{fig:simulation:orch:fault}
    \end{subfigure}
    \hfill{}
    \begin{subfigure}[b]{0.23\textwidth}
        \centering
        \includegraphics[width=\linewidth]{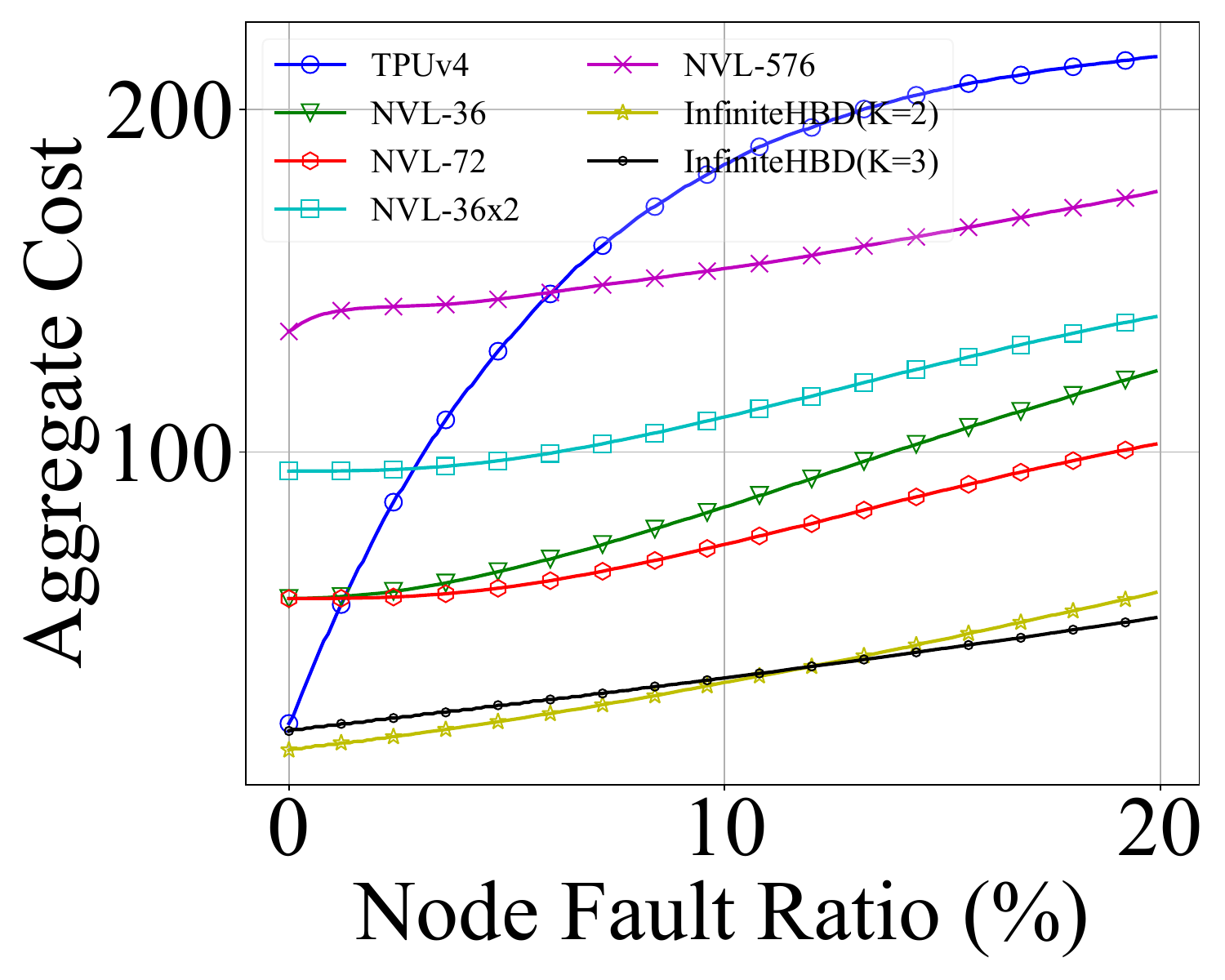}
        \caption{Aggregate cost.}
        \label{fig:eval:aggregate-cost}
    \end{subfigure}
    \vspace{-10pt}
    \caption{DCN traffic optimization analysis and aggregate normalized cost varies across different architectures under different fault ratios.}
    \label{fig:simulation:job_scale:orch}
    \vspace{-10pt}
\end{figure*}

This section examines the impact of orchestration algorithms on DCN communication efficiency. Experiments were performed on a Fat-Tree architecture, like the setup in ~\cite{sigcomm2024meta}. As shown in \figref{fig:simulation:orch:cluster}, the algorithm is not sensitive to cluster size. Therefore, the evaluation is based on TP-32 operations on \sys{} with 8192 GPUs. 

\begin{itemize}[itemsep=2pt,topsep=0pt,parsep=0pt, leftmargin=2ex]
    \item \textbf{Baseline:} A greedy algorithm, which randomly selects nodes from the cluster and uses the first permutation that meets the requirements.
    \item \textbf{Optimized:} The HBD-DCN orchestration algorithm proposed in \secref{sec:design:orch}.
\end{itemize}  

\figref{fig:simulation:orch:job} illustrates the impact of job-scale ratios (job size/total cluster GPUs) on cross-ToR traffic, where node fault ratio is 5\%. The Baseline consistently results in approximately 10\% cross-ToR traffic. In contrast, the Optimized algorithm significantly outperforms the Baseline, reducing cross-ToR traffic to just 1.72\% even at a 90\% job-scale ratio.
\figref{fig:simulation:orch:fault} explores the sensitivity to node faults, with the job scale ratio fixed at 85\%. The Baseline shows a linear increase of cross-ToR traffic, while the Optimized algorithm sustains near-zero cross-ToR traffic for fault ratios under 7\%.

\subsection{Cost and Power Analysis}
\label{sec:simulation:cost-power}

To evaluate the interconnect costs of HBD architectures, we gather the cost and power information with the following methodologies:

\begin{itemize}[itemsep=2pt,topsep=0pt,parsep=0pt, leftmargin=2ex]
    \item For standard components (DAC cables, optical transceivers, fibers), pricing is sourced from official retailer websites~\cite{FS_COM, FIBER_MALL, NADDOD} with a 60\% wholesale discount validated against internal data.
    \item For components with scarce public pricing information, such as Google Palomar OCS, NVIDIA NVLink Switch, 1.6 Tbps ACC cables/optical transceivers, the data is amalgamated from multiple sources~\cite{SEMIANALYSIS_GB200, SEMIANALYSIS_OCS, SEMIANALYSIS_Power} to enhance accuracy.
    \item Public power consumption data is available for most components, though for NVLink Switch, multiple sources are combined to estimate a reasonable value.
\end{itemize}

The breakdown analysis of each architecture is provided in the Appendix~\S\ref{appendix:cost}. Based on this, the cost and power consumption are normalized according to GPU count and per-GPU bandwidth. As depicted in \tabref{tab:eval:cost-power}, \sys{} exhibits the lowest interconnect cost per GPU per GBps. Under the $K =2$ configuration, its cost is only 62.84\% of Google TPUv4 and 30.86\% of the NVIDIA GB200 NVL-36/72, with minimal power consumption.
This efficiency is primarily attributed to the avoidance of centralized switches. TPUv4 ranks second in interconnect cost and lowest in power consumption, achieved by reducing optical module use and per-port OCS costs. The NVL series has higher interconnect costs and power consumption due to its fully-connected topology and high-cost NVLink Switches. Notably, NVL-576 incurs the highest cost and power consumption due to its multilayer non-blocking topology, which increases optical module expenses and requires more NVLink Switches.

\begin{table}[h!t] \small
    \centering
    \begin{tabular}{lcccc}
    \toprule
    \textbf{Architecture}  & \multicolumn{2}{c}{\textbf{Per-GPU}}  & \multicolumn{2}{c}{\textbf{Per-GPU Per-GBps}} \\
 &  Cost & Watts & Cost & Watts \\
    \midrule
    TPUv4  & 1567.20  & 19.39 & 5.22& 0.06 \\
    NVL-36  & 9563.20  & 75.95 & 10.63& 0.08 \\
    NVL-72  & 9563.20  & 75.95 & 10.63 & 0.08 \\
    NVL-36x2  & 17924.00  & 150.33 & 19.92  & 0.17\\
    NVL-576   & 30417.60  & 413.45 & 33.80  & 0.46\\
    \midrule
    \SYS{} ($K=2$) &  2626.80 &  48.10 & 3.28  & 0.06\\\
    \SYS{} ($K=3$) &  3740.60 &  72.05  & 4.68  & 0.09\\
    \bottomrule
    \end{tabular}
    \caption{Interconnect cost (\$) and power ($Watts$).}
    \label{tab:eval:cost-power}
    \vspace{-2em}
\end{table}

Beyond interconnect costs, fault resilience variations also affect aggregate costs. The aggregate cost is defined as:

$$Cost_{GPU} \times (N_{Wasted-GPU} + N_{Faulty-GPU}) + Cost_{Interconnect}$$

Simulations on a 3K-GPU cluster using the TP-32 configuration evaluate GPU availability under varying fault ratios across different architectures.
The variation in aggregate cost for different HBD architectures under varying node fault ratios is illustrated in \figref{fig:eval:aggregate-cost}. \sys{} consistently exhibits the lowest aggregate cost. Furthermore, when the fault ratio is below 12.1\%, the aggregate cost of \sys{} ($K=2$) is less than that of \sys{} ($K = 3$), suggesting that ($K = 2$) is the optimal design for most scenarios.

\section{Discussion}

\para{AllToAll communication.}
Ring topology in \sys{} struggles with AllToAll communication (e.g., EP), exhibiting poor performance at $O(p^2)$, where $p$ is the group size. This can be improved by linking backup lines to nodes indexed at $n\pm 2^i$ instead of $n\pm i$ and applying the Binary Exchange algorithm, reducing time complexity to $O(p\log_2 p)$. During the algorithm, \ocstrx{} needs to connect to different GPUs with runtime switching; since \ocstrx{} switches in 60-80 $\mu s$, reconfiguration can be overlapped with computation. For $K=2$ \sys{} designs, performance matches the ideal Bruck algorithm~\cite{bruck} when $p<8$. However, this design introduces complexities in construction, failover, and orchestration, and necessitates GPU routing capabilities. Therefore, it is not applied.
\revised{We provide a detailed theoretical discussion on how \sys{} supports AllToAll communication in Appendix~\S\ref{appendix:all2all}.}

\para{Simulation Scale.} 
Simulations using real fault traces from 3,200 GPUs, as detailed in \secref{sec:simulation:fault}, were conducted on a cluster comprising 2,880 GPUs. This is because the simulation's GPU count must be less than the total GPUs in the fault trace, and 2,880 is the largest number divisible by 576 and less than 3,200. This configuration allows the entire cluster to be divided into 5 NVL-576 units for the simulation.

\para{Multi-dimension parallelism.}
\sys{} is optimized for single-dimension parallelism. To support multi-dimensional communication, two approaches are viable. 1) \textit{Independent Interconnects: } Each \ocstrx{} bundle includes multiple \ocstrx{} units (e.g., 4 or 8), then links each of the units to a separate inter-host topology. This isolates parallel dimensions but results in fixed bandwidth per dimension, leading to inefficiencies. 2) \textit{Time-Division multi-dimension: } Main and backup lines of \ocstrx{} can be used to form separated inter-host topologies. Rapidly switching between them can support multi-dimensional parallelism. However, this introduces complexity in managing multi-dimensional overlap and reduces the fault tolerance of \sys{}.

\para{Single-Job vs. Multi-Job.} 
Existing studies explore multi-job scheduling in GPU clusters~\cite{mlass,lyra}. Deploying certain small jobs, such as inference tasks, can mitigate GPU fragmentation. However, given the shortage of GPUs in LLM training, any idle GPU—whether repurposed for small jobs or not—is undesirable. Thus, \sys{} prioritizes single-job execution for simplicity. 

\para{OCS vs. EPS.} 
\ocstrx{} enables multi-path selection, a feature also achievable with Electronic Packet Switching (EPS). For example, the inter-host topology of \sys{} can be implemented using UBB 2.0-based servers by adding external optical interfaces to switches in servers. However, this would require twice the number of optical modules and numerous high-throughput switching chips for the entire system, significantly increasing cost and power consumption compared to \ocstrx{}.

\section{Related Work}

\para{HBD Architectures.}  
HBDs are crucial for enabling communication-intensive parallelism strategies (TP/EP) for LLM training. NVIDIA DGX SuperPOD~\cite{superpod} and GB200 NVL series~\cite{nvl72} use any-to-any electrical switching, delivering high performance but suffering from high costs, scalability limitations, and fragmentation. In contrast, direct interconnect HBDs like Dojo~\cite{dojo}, TPUv3~\cite{cacm2020tpuv3}, and SiP-Ring~\cite{sip-ml} improve scalability but have a large fault explosion radius. TPUv4~\cite{isca2023tpu} and TPUv5p~\cite{tpuv5} attempt a middle ground but still lack full node-level fault isolation. \sys{} introduces a novel architecture that reduces cost, improves scalability, minimizes fragmentation, enhances fault isolation, and dynamically supports TP.  

\para{AI DCN Architectures.}  
MegaScale~\cite{megascale} and Meta’s~\cite{sigcomm2024meta} AI DC use Clos-based topologies, while Rail-Optimized~\cite{rail-optimized} and Rail-Only~\cite{wang2024railonly} architectures optimize for LLM traffic patterns. Alibaba HPN~\cite{sigcomm2024hpn} enhances fault tolerance with a dual-plane design. \sys{} is compatible with all of them for LLM training.  

\para{OCS Technologies.}  
OCS enables dynamic topology reconfiguration in datacenters~\cite{missionapollo, isca2023tpu, mfabric}. A MEMS OCS-based switch supports high port counts~\cite{missionapollo, mems-320}, while Silicon Photonics (SiPh) achieves lower reconfiguration latency and cost~\cite{thermo-optic_2006}. This work proposes a SiPh-based OCS transceiver (\docs), which constructs an interconnect fabric without centralized switches.  

\para{Reconfigurable Networks.}  
Traditional studies~\cite{helios,c-through,osa,mordia,sirius,xia2015enabling,megaswitch,rotornet,opera,firefly,shale} focus on generic DCN architectures without optimizing for LLM training traffic, leading to suboptimal topologies. Recent advancements like SiP-ML~\cite{sip-ml}, TopoOpt~\cite{topoopt2023}, and mFabric~\cite{mfabric} introduce dedicated training optimizations but still underutilize optical network reconfigurability for better fault tolerance and GPU utilization.  

\para{AI Job Schedulers.}  
Schedulers such as ~\cite{gandiva,themis,tiresias, {byteps_1}, {byteps_2}, pollux} aim to improve GPU utilization. However, they exhibit dual limitations: their designs are premised on a non-reconfigurable network, while also failing to consider job scheduling within HBD for optimizing traffic patterns in DCN. This work proposes an HBD-DCN orchestration algorithm based on reconfigurable networks to address these limitations.

\section{Conclusion}

In this paper, we propose \sys{}, a novel HBD design that supports datacenter scale, dynamic TP group size and near-ideal fault explosion radius.
\sys{} is built upon a design of optical transceivers integrated with SiPh-based \ocstrx{}, a reconfigurable K-Hop Ring topology and an HBD-DCN orchestration algorithm to leverage the capabilities of the new hardware. 
Using a real fault trace of 3K-GPU cluster and the in-house simulator, we demonstrate that \sys{} achieves GPU utilization close to the ideal model during faults, delivers superior cost and energy efficiency compared to existing designs, and provides effective control over cross-ToR DCN traffic.
We believe \sys{} provides an efficient scaling solution for HBD, which offers new insights for the next-generation infrastructure for training trillion-parameter LLMs.

\begin{acks}
    We thank our shepherd Yiting Xia and anonymous reviewers for their constructive feedback. We are grateful to Juston Zhu, Zhijie Chen, Ziyao Shen, Jie Chen, and Edmund Xu from Lightelligence Pte. Ltd. for their assistance in hardware evaluation. We also thank Instruments and Electronics (Shanghai) Associates (INESA) for providing a GPU cluster optical interconnect testing platform. This work is supported by Beijing Municipal Science and Technology Project No. Z241100004224023, National Natural Science Fund for the Excellent Young Scientists Fund Program (Overseas), and Peking University startup fund.
\end{acks}

\clearpage

\newpage
\bibliographystyle{ACM-Reference-Format}
\bibliography{paper}

\newpage
\clearpage
\begin{appendices}

\section*{APPENDICES}
Appendices are supporting material that has not been peer-reviewed.

\section{Production Fault Trace}
\label{appendix:production-fault-trace}
\begin{figure}[h!t]
    \centering
    \begin{subfigure}[b]{0.23\textwidth}
        \centering
        \includegraphics[width=\textwidth]{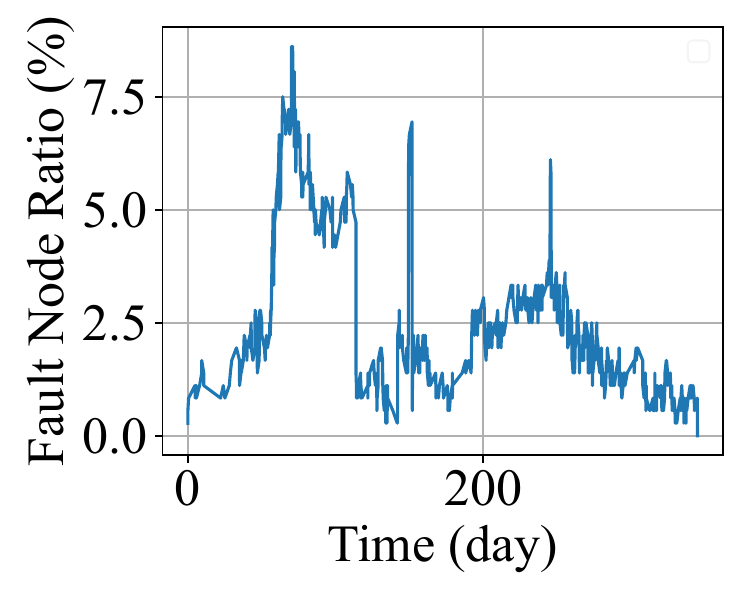}
        \caption{Fault Node Ratio Trace.}
        \label{fig:simulation:trace:timetrace}
    \end{subfigure}
    \hspace{2pt}
    \begin{subfigure}[b]{0.23\textwidth}
        \centering
        \includegraphics[width=\textwidth]{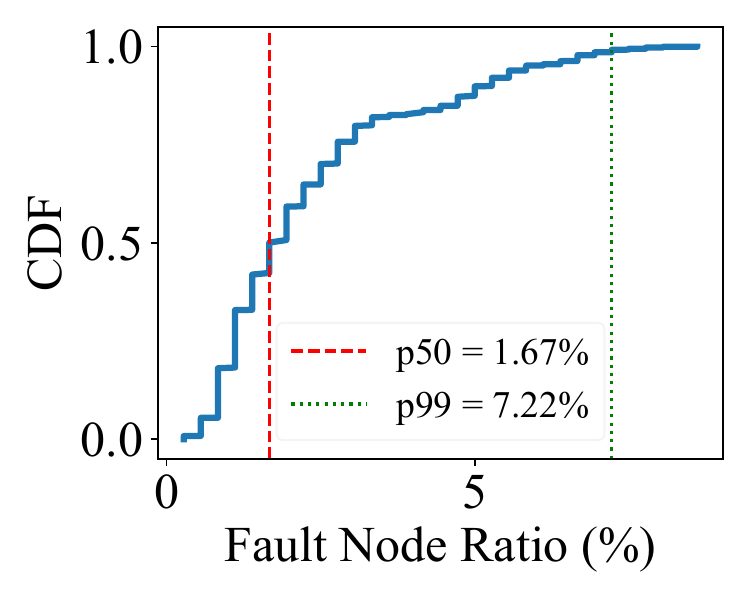}
        \caption{Cumulative Distribution.}
        \label{fig:simulation:trace:cdf}
    \end{subfigure}
    \vspace{-2ex}
    \caption{Fault node trace in the production AI DC.}
    \label{fig:simulation:trace}
\end{figure}

The production fault trace was collected from a 3K-GPU cluster dedicated to pretrain with 8-GPU nodes during a period of 348 days. The trace includes details such as fault start time, fault end time, and the ID of the faulty node. \figref{fig:simulation:trace:timetrace} and \figref{fig:simulation:trace:cdf} provide a macro-level overview of the production fault trace. On average, the ratio of faulty 8-GPU nodes at any given time is $2.33\%$, with a P99 value of $7.22\%$.

Since most of the failure events are GPU faults, we normalized the trace of 8-GPU nodes to generate 4-GPU nodes trace. Assuming that the fault rates of GPUs are i.i.d. with a fault probability of $p$ for each GPU, and considering that a node is deemed faulty if any GPU within it fails, the fault rate of an 8-GPU node is calculated as:  

\vspace{-1em}
$$
P_{fault}(8\text{-GPU}) = 1 - (1-p)^8 = 2.33\%.
$$  

From this, we derive $p = 0.29\%$. The fault rate for a 4-GPU node is then:  
$$
P_{fault}(4\text{-GPU}) = 1 - (1-p)^4 = 1.17\%.
$$  

The fault event trace for 4-GPU nodes is generated using Bayes’ theorem as follows:

\begin{align*}\label{eq:convert-trace}
& P_{fault}( \text{4-GPU} \mid  \text{8-GPU})\\ 
    &=\frac{P_{fault}(\text{8-GPU} \mid \text{4-GPU}) P_{fault}(\text{4-GPU})}{P_{fault}(\text{8-GPU})} \\ 
    & =  \frac{1 \times 1.17\%}{2.33\%} = 50.21\% \\
\end{align*}

Thus, whenever a fault occurs in an 8-GPU node in the original trace, each of the two corresponding 4-GPU nodes at the same location has a $50.21\%$ probability of failure. This method is used to convert the traces.

As node faults are assumed to be i.i.d., the simulator linearly maps the fault trace onto different network architectures.

\section{GPT-MoE Architecture}
\label{appendix:gpt-moe}
This model is a mixture-of-experts (MoE) model with the following configuration:

\para{Model Configuration:}
\begin{itemize}
    \item \textbf{Number of Layers:} 192
    \item \textbf{Inner Layer Dimension:} 49152
    \item \textbf{Embedding Dimension:} 12288
    \item \textbf{Hidden Dimension:} 12288
    \item \textbf{Vocabulary Size:} 64000
    \item \textbf{Number of Attention Heads:} 128
    \item \textbf{Maximum Sequence Length:} 2048
    \item \textbf{Number of Experts:} 8
    \item \textbf{MoE Layer Ratio:} 0.5
    \item \textbf{Top-K Experts:} 2
\end{itemize}

\para{Runtime Configuration:}
\begin{itemize}
    \item \textbf{Virtual Pipeline Parallelism:} 3
    \item \textbf{Micro Batch Size:} 1
    \item \textbf{Global Batch Size:} 1536
    \item \textbf{Max Sequence Length:} 2048
\end{itemize}

\section{Theoretical analysis of GPU waste ratio for \sys}
\label{appendix:ft-anay}

The number of backup lines, given by $2K - 2$, will significantly influence the fault tolerance of \sys. We use the expectation of waste ratio caused by GPU failure and the fragmentation problem to evaluate this design, the result is shown in \tabref{table:design:1.5ratio}.

For a single working node in the middle of the line, the count of breakpoints $B$ on its two sides has the expectation as:

\vspace{-1em}
\begin{equation*}
E_B(\eta = 1,middle) = 2(P_s^K + P_s^{2K})
\end{equation*}

where $P_s$ is the failure probability of GPU node, and $\eta$ is the count of nodes. 

Once the distance between one node and the tail of the line is $\alpha < K$, it will connect to all nodes between itself and the last one, so there will be no breakpoints on this side, and the expectation of breakpoints count is less than nodes in the middle of line.
Then, for any node in the line topology:

\vspace{-1em}
$$
E_B(\eta = 1) \leq E_B(\eta = 1,middle) 
$$

When the distance between two nodes is $\beta \geq K$, the breakpoints among them can be considered independent.
Once the distance $\beta < K$, as all nodes in this range are connected to these two nodes, there will be no breakpoints between them. So, the expectation is less than two independent nodes. Then,

\vspace{-1em}
\begin{align*}
E_B(\eta =& 2) < E_B(\eta = 2, \beta \geq K) =  2E(\eta = 1)   \\ 
 E_B(\eta =& N_s) \leq N_s E_B(\eta = 1) 
\end{align*}

\begin{figure}[h!t]
    \centering
    \includegraphics[width=0.8\linewidth]{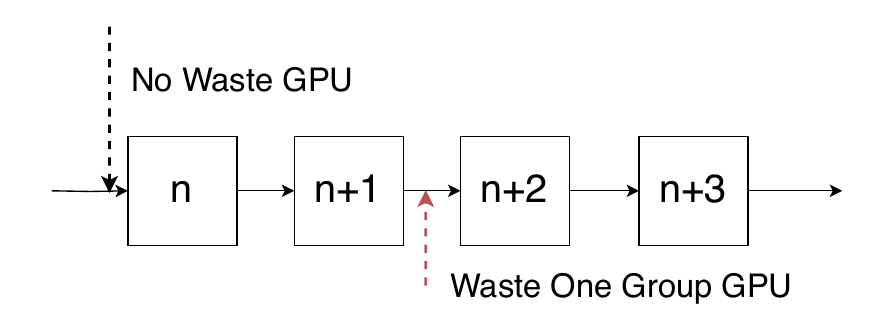}
    \caption{Break point can cause node waste compared to ideal situation.}
    \vspace{-1em}
    \label{fig:subline-waste}
\end{figure}

For a LLM job which require a ring communication size (TP .etc) as $N_t$, \sys   will cut the whole line topology into several sub-lines with the length of $N_t/R$.
Once \sys cuts a new sub-line from the remaining nodes in the line, 
all $N_t$ GPUs will be wasted when one break point exists in the middle of this sub-line required, shown in \fig{fig:subline-waste}. 
Then the expectation for waste GPU caused by one single break point is:

\vspace{-1em}
$$
E_W(B=1) = N_t R\cdot (1 - (N_t/R)^{-1} ) = R(N_t -R)
$$

As the influence between two break points only reduce the expectation of wasted GPUs, we can have this for $X$ break points:

\vspace{-1em}
\begin{equation*}
E_W(B = X) \leq XE_W(B=1) = XR(N_t-R)
\end{equation*}

So the expectation of wasted GPU for a cluster with $N_s$ GPU nodes is:

\vspace{-1em}
\begin{align*}
E_W(\eta = N_s) &\leq \sum P(B=X ,\eta = N_s) \cdot X\cdot  E_W(B=1)\\
&= E_B(\eta = N_s)\cdot E_W(B=1)\\
&\leq  \lim_{P_s\rightarrow 0}2N_s\cdot R \cdot (N_t-R)P_s^K
\end{align*}

The final expectation of GPUs waste ratio is \eqref{eq:design:ratio}:

\begin{equation}
E_{WR}(\eta = N_s) = \frac{E_W(\eta = N_s)}{N_g} \leq 2(N_t-R)(P_s)^K
\label{eq:design:ratio}
\end{equation}

In our trace from a 3K-GPU cluster over 348 days, the 99th percentile (p99) failure rate for 8-GPU nodes is 7.22\%. Therefore, we set the node failure rate $P_s = 7.22\%$ when $R = 8$. Based on the calculations in Appendix~\S\ref{appendix:production-fault-trace} , given a GPU failure rate of $0.93\%$, we derive the corresponding node failure rate $P_s = 3.67\%$ for $R = 4$. If a TP32 job is running on \sys{} (i.e., $N_t = 32$), plugging these values into Equation~\eqref{eq:design:ratio}, we obtain the upper bound of the expected waste ratio for different configurations, as shown in \tabref{table:design:1.5ratio}.

\begin{table}[h!t]
\centering
\begin{tabular}{cccc}
    \toprule
        & $K=2$&$K=3$&$K=4$\\
    \midrule
     R=4& $7.54\%$ & $0.28\%$ & $1.02\times 10^{-4}$ \\
     R=8& $25.02\%$ & $1.81\%$ & $0.13\%$ \\
     \bottomrule
\end{tabular}
\caption{Upper bound for waste ratio expectation of GPU, where GPU failure rate is 0.93\% and $N_t$ is 32}
\vspace{-2em}
\label{table:design:1.5ratio}
\end{table}

For example, the 4-GPU node ($R = 4$), 3-bundle ($K = 3$) design incurs less than 0.28\% additional GPU waste, while the waste ratio for the $R = 8, K = 4$ configuration remains below 0.13\%. Both are sufficiently low for production cluster deployment.

\section{Orchestration For Fat-Tree}
\label{appendix:orch-algo}
In this section, we introduce the orchestration algorithm under Fat-Tree DCN in detail.

\para{Notations}
\label{appendix:orch-algo:notation}
To ensure rigorous mathematical reasoning, we introduce the following notations:

\begin{itemize}
    \item {
        $n$: number of nodes in the data-center.
    }
    \item {
        $K$: \docs{} bundle (see \S\ref{section:design:topology}).
    }
    \item {
        $S_{all}$: ordered set, represents all nodes numbered from 1 according to their physical connection order in DCN fabric. $|S_{all}|=n$.
    }
    \item {
        $S$: ordered subset, represents nodes, $\forall u \in S, u \in S_{all}$. Adjacent elements in $S$ are also adjacent from the perspective of the \SYS{} topology. 
    }
    \item{
        $E$: The set of edges across $S$, should be equal to $\{ (S_i, S_j) \mid 1 \leq i < j \leq n, j - i \leq K \} $, representing the connections between nodes, including both primary and backup links, and $O(|E|) = O(K|S|)$.
    }
    \item {
        $InfHBD=<S,E>$: the topology of \SYS{} as an undirected graph.
    }
    \item {
        $F$: faulty nodes.
    }
    \item {
        $HealthyHBD=<H,HE>$: healthy node subgraph where the set of healthy nodes $H = S - F$ and the edge set $HE = \{ (u, v) \mid u \in H \text{ and } v \in H \text{ and } (u, v) \in E \}$.
    }
    \item{
        $t$: TP size, number of GPUs in one TP Group.
    }
    \item{
        $r$: GPU ranks per node.
    }
    \item{
        $m=t/r$: number of nodes in a TP group.
    }
    \item{
        $s$: job scale, number of GPUs required for the job.
    }
    \item{
        $d$: Aggregation-Switches Domain size. Number of nodes under coverage of one group of Aggregation-Switches.
    }
    \item{
        $n_{constrains}$: number of applied constraints in binary-search-based orchestration algorithm.
    }
    \item{
        $p$: number of nodes under each ToR.
    }
    \item{
        $l$: shortest sub-line length under fat-tree orchestration.
    }
    \item{
        $n_{maxsubline}=\lfloor \frac{nd}{p} \rfloor$: max number of sub-lines.
    }
    \item{
        $G_{deploy}=<S_{deploy},E_{deploy}>$: deployed topology. After applying the deployment strategy, the topology from the perspective of \SYS{} is described as follows: $S_{\text{deploy}}$ is an ordered set where adjacent elements correspond to adjacent nodes in \SYS{}, and $E_{\text{deploy}}$ represents the connections between nodes.
    }
    
\end{itemize}

The orchestration algorithm (\algref{alg:orchestration-ideal}) without considering DCN has the overall time complexity $3\cdot O(|H| + |HE|) = O(|S| + |E|) = O((K+1)|S|) = O(|S|)$.

\begin{algorithm}[!h]
\small
\caption{Orchestration-DCN-Free}
\label{alg:orchestration-ideal}
\SetAlgoNlRelativeSize{-1}
\SetAlgoNlRelativeSize{1}
\KwIn{$\text{InfHBD}=\langle S, E \rangle$, $F$, $m$}
\KwOut{ Placement scheme that maximizes GPU utilization}

 Initialize $H = S - F$\;
 Initialize $HE = \{ (u, v) \mid u \in H \text{ and } v \in H \text{ and } (u, v) \in E \}$\;
 Create subgraph $HealthyHBD = \langle H, HE \rangle$\;
 Initialize $component\_list = []$\;
 Initialize $visited = \{\}$\;
 Initialize $placement\_scheme= \{\}$\;

\For{ each node $s$ in $H$}
{
    \uIf{ $s$ not in $visited$}
    {
         $component = Connected-Component-DFS(s, HealthyHBD, visited)$\;
         Add $component.sortedInHBD()$ to $component\_list$\;
    }
}
\For{ each $component$ in $component\_list$}
{
    \While{ $component.size()\geq m$}
    {
         Add $component.pop(m)$ to $placement\_scheme$\;
    }
}
        
 \KwRet{$placement\_scheme$}
 \end{algorithm}

\begin{algorithm}[!h]
\small
\caption{Deployment-Strategy}
\label{alg:deployment-strategy}
\SetAlgoNlRelativeSize{-1}
\SetAlgoNlRelativeSize{1}
 \KwIn{Node ordered set $S$, \docs{} direction $K$, parallel factor $p$}
 \KwOut{Deployment topology $G_{deploy}=<S_{deploy},E_{deploy}>$}
 Initialize ordered set $S_{deploy}=[]$\;
 Initialize $l=\lfloor \frac{|S|}{p}\rfloor$\;
\For{$i$ in $0...p-1$}
{
    \For{$j$ in $0...l-1$}{
         Add $i+j\cdot p$ to $S_{deploy}$\;}
}
 Create $E_{deploy}=\{(S_{deploy}^i,S_{deploy}^j)|1\leq i\le j\leq |S_{deploy}|, j-i\leq K \}$\;
 \KwRet{$G_{deploy}=<S_{deploy},E_{deploy}>$}
\end{algorithm}

The Fat-Tree is another common topology used in data centers. A typical training strategy for this topology aims to maximize the bandwidth utilization under ToR (Top of Rack) Switches. Using Meta's two-stage clos topology\cite{sigcomm2024meta} as a reference, it can be observed that there is an attempt to run CP under ToR Switches.

\para{Deployment Strategy.} Assuming there are $p$ nodes under each ToR, nodes with the same index under each ToR are deployed along the same parallel sub-line, and the $p$ sub-lines are connected end-to-end, as shown in \fig{fig:fat-tree-topo}. The training strategy involves running CP $p$ across the sub-lines and running TP within them.

\para{Orchestration Constraints.} To maximize the utilization of ToR bandwidth and minimize cross-ToR traffic, the fat-tree topology introduces two constraints:

\begin{packeditemize}
    \item {
        \textbf{Aggregation-Switches Domain Constraint:} The coverage domain of a group of Aggregation Switches is limited, meaning that TP groups spanning across Aggregation Switches domains would result in cross-rail traffic, which should be avoided as much as possible.
    }
    \item {
        \textbf{TP Group Alignment Constraint:} A CP Group consists of TP Groups across parallel sub-lines. To keep CP traffic within the ToR switches, the TP Groups must be aligned. If a node fails under one ToR, all nodes under that ToR are considered to have failed, expanding the failure radius by a factor of $p$. 
    }
\end{packeditemize}

\para{Binary-Search-Based Orchestration Algorithm.} Based on the constraints and deployment strategy, we develop a binary search orchestration algorithm (see \algref{alg:orchestration-fat-tree}) that adjusts the number of satisfied constraints. The binary search first relaxes the TP Group alignment constraints within the Aggregation-Switches Domain and then relaxes the TP Group crossing constraints between Aggregation-Switch domains (see \algref{alg:placement-fat-tree}). This process is monotonic.

The time complexity of \algref{alg:orchestration-ideal} is $O(|S|)$, and the complexity of \algref{alg:placement-fat-tree} is:  

$$\sum_{i=1}^{n_{subline}} O(|S_{subline}|) = O(\sum_{i=1}^{n_{subline}} |S_{subline}|) = O(|S_{all}|) = O(n)$$  

Thus, the overall time complexity of \algref{alg:orchestration-fat-tree} is $O(n \log n)$.

\begin{algorithm}[!h]
\small
\caption{Placement-Fat-Tree}
\label{alg:placement-fat-tree}
\SetAlgoNlRelativeSize{-1}
\SetAlgoNlRelativeSize{1}
 \KwIn{$G_{deploy}=<S_{deploy},E_{deploy}>$, $n_{constraints}$, $F$, $l$, $m$, $n_{maxsubline}$, $d$, $p$}
 \KwOut{Placement scheme}
 Initialize $placement\_scheme=\{\}$\;
 Initialize $n_{align}=max(0,n_{constraints}-n_{maxsubline})$, $n_{subline}=min(n_{maxsubline},n_{constraints})$\;
 
\For{$i$ in $0..n_{align}-1$}
{
    \For{$j$ in $1..d$}
    {
        $sid=i*d+j$\;
        \If{$sid \in F$}
        {
            $F\cup \{\lfloor \frac{sid-1}{p}\rfloor\cdot p+1..(\lfloor \frac{sid-1}{p}\rfloor+1)\cdot p \}$\;
        }
    }
}
\For{$i$ in $1..n_{subline}$}
{
     $S_{subline}=S_{deploy}.pop(l)$\;
     $E_{subline}=\{(u,v)\mid u\in S_{subline} \text{ and } v\in S_{subline} \text{ and } (u,v)\in E_{subline}\}$\;
     $F_{subline}=F\cap S_{subline}$\;
     $placement\_scheme=placement\_scheme\cup \text{Orchestration-Ideal}(<S_{subline},E_{subline}>, F_{subline}, m)$\;
}
 $E_{res}=\{(u,v)\mid u \in S_{deploy} \text{ and } v \in S_{deploy} \text{ and } (u,v) \in E_{deploy}\}$\;
 $F_{res}=F\cap S_{deploy}$\;
 $placement\_scheme=placement\_scheme\cup \text{Orchestration-Ideal}(<S_{deploy},E_{res}>, F_{res},m)$\;
 \KwRet{$placement\_scheme$}
\end{algorithm}

\begin{algorithm}[!h]
\small
\caption{Orchestration-Fat-Tree}
\label{alg:orchestration-fat-tree}
\SetAlgoNlRelativeSize{-1}
\SetAlgoNlRelativeSize{1}
 \KwIn{$S$, $r$, $p$, $F$, $t$, $s$, $d$, $K$.}
 \KwOut{Placement scheme that satisfies job scale and minimizes cross-rail traffic.}
 Initialize $m=t/r$, $n=|S|$, $l=\lfloor\frac{d}{p}\rfloor$\, $n_{domain}=\lfloor\frac{n}{d}\rfloor$, $n_{maxsubline}=\lfloor\frac{nd}{p}\rfloor$\;
 Create graph $G_{deploy}=<S_{deploy},E_{deploy}>=\text{Deployment-Strategy}(S,K,p)$\;
 Initialize $high=n_{domain}+n_{maxsubline}$\;
 Initialize $low=0$\;
 Initialize $placement\_scheme=\{\}$\;
\While{ $low \leq$ high}
{
     $mid=\lfloor \frac{low+high}{2} \rfloor$\;
     $placement\_scheme=\text{Placement-Fat-Tree}(G_{deploy},mid,F,l,m,n_{maxsubline},d,p)$\;
    \eIf {$|placement\_scheme|\cdot m\cdot r\ge s$}
    {
         $low=mid+1$\;
    }
    {
         $high=mid-1$\;
    }
}
    
\eIf{$|placement\_scheme|\cdot m\cdot r\ge s$}
{
    \KwRet {$placement\_scheme$}
}
{
    \KwRet {None}
}
\end{algorithm}

\section{Additional Simulation Results for Fault Resilience}
\label{appendix:wasted-GPUs-ratio}

\begin{figure*}[h!t]
    \centering
    \begin{subfigure}[b]{0.23\linewidth}
        \centering
        \includegraphics[width=\linewidth]{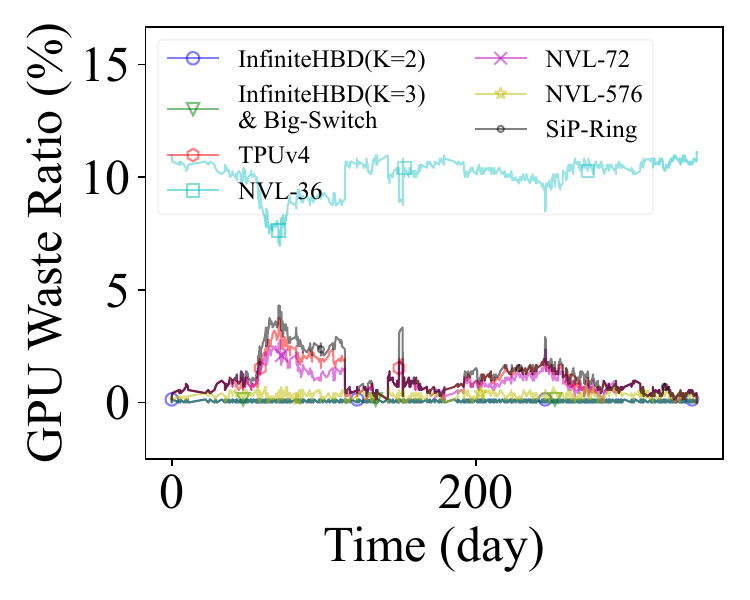}
        \caption{TP-8.}
        \label{fig:simulation:wasted-trace:tp8-4gpu}
    \end{subfigure}
    \hspace{2pt}
    \begin{subfigure}[b]{0.23\linewidth}
        \centering
        \includegraphics[width=\linewidth]{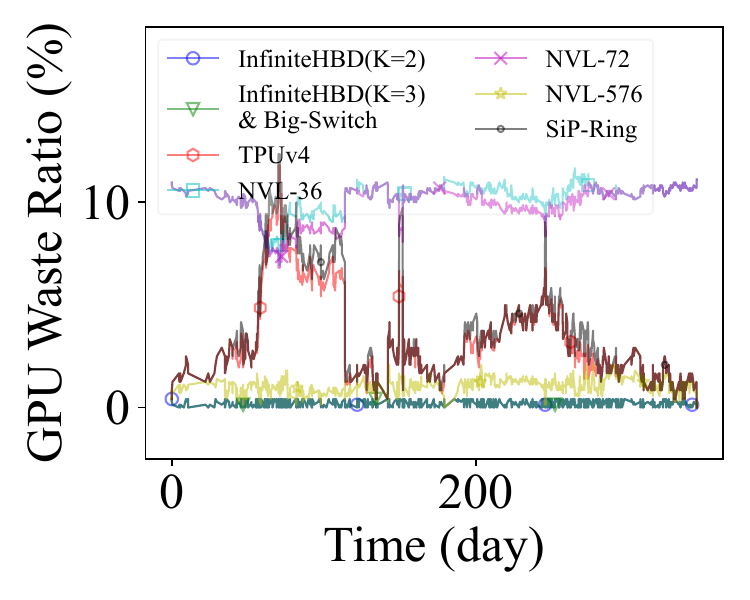}
        \caption{TP-16.}
        \label{fig:simulation:wasted-trace:tp16-4gpu}
    \end{subfigure}
    \hspace{2pt}
    \begin{subfigure}[b]{0.23\linewidth}
        \centering
        \includegraphics[width=\linewidth]{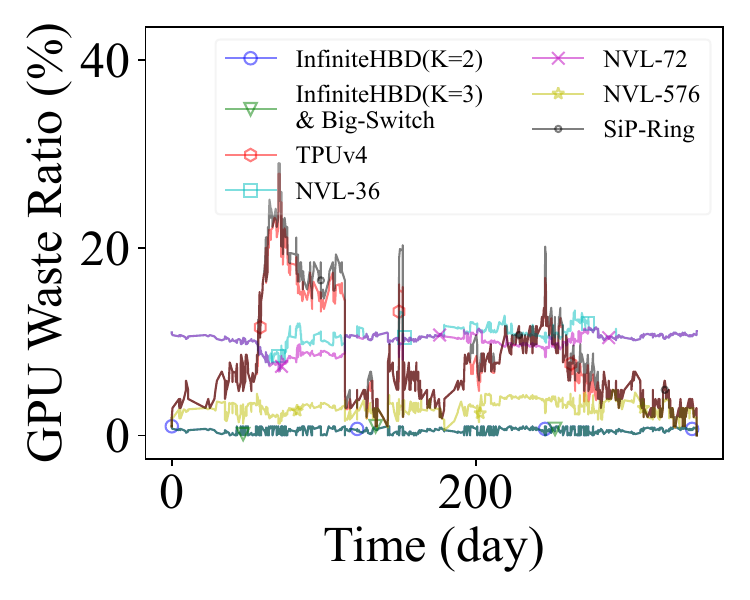}
        \caption{TP-32.}
        \label{fig:simulation:wasted-trace:tp32-4gpu}
    \end{subfigure}
    \hspace{2pt}
    \begin{subfigure}[b]{0.23\linewidth}
        \centering
        \includegraphics[width=\linewidth]{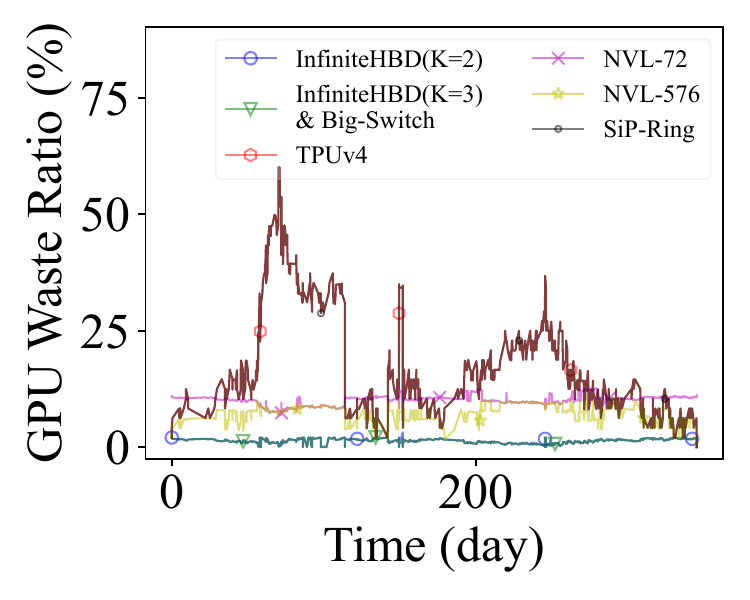}
        \caption{TP-64.}
        \label{fig:simulation:wasted-trace:tp64-4gpu}
    \end{subfigure}

    \vspace{-1ex}
    \caption{GPU waste ratio over production fault trace, 4-GPU node.}
    \label{fig:simulation:wasted-trace}
\end{figure*}

\begin{figure*}[h!t]
    \centering
    \begin{subfigure}[b]{0.23\linewidth}
        \centering
        \includegraphics[width=\linewidth]{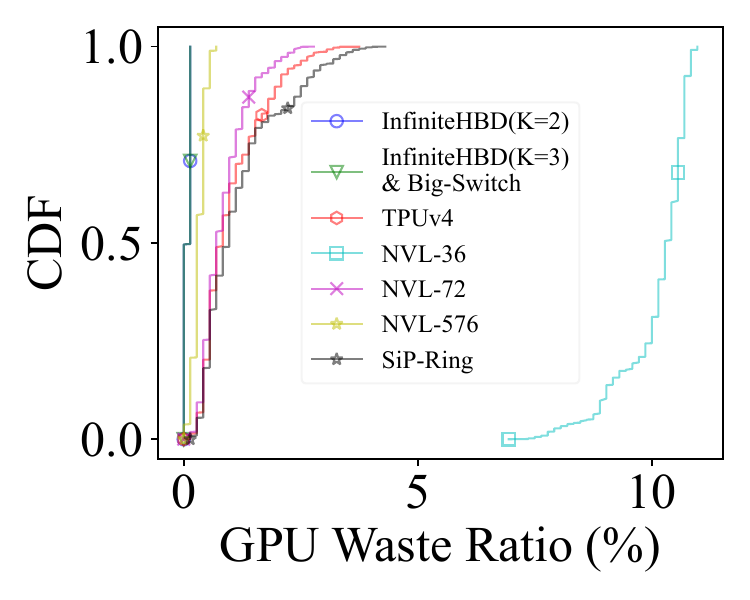}
        \caption{TP-8.}
        \label{fig:simulation:waste-cdf:tp8-gr4}
    \end{subfigure}
    \hspace{2pt}
    \begin{subfigure}[b]{0.23\linewidth}
        \centering
        \includegraphics[width=\linewidth]{figs/evaluation/fault_trace_based/cdf_trace_waste_tp16_gr4.pdf}
        \caption{TP-16.}
        \label{fig:simulation:waste-cdf:tp16-gr4}
    \end{subfigure}
    \hspace{2pt}
    \begin{subfigure}[b]{0.23\linewidth}
        \centering
        \includegraphics[width=\linewidth]{figs/evaluation/fault_trace_based/cdf_trace_waste_tp32_gr4.pdf}
        \caption{TP-32.}
        \label{fig:simulation:waste-cdf:tp32-gr4}
    \end{subfigure}
    \hspace{2pt}
    \begin{subfigure}[b]{0.23\linewidth}
        \centering
        \includegraphics[width=\linewidth]{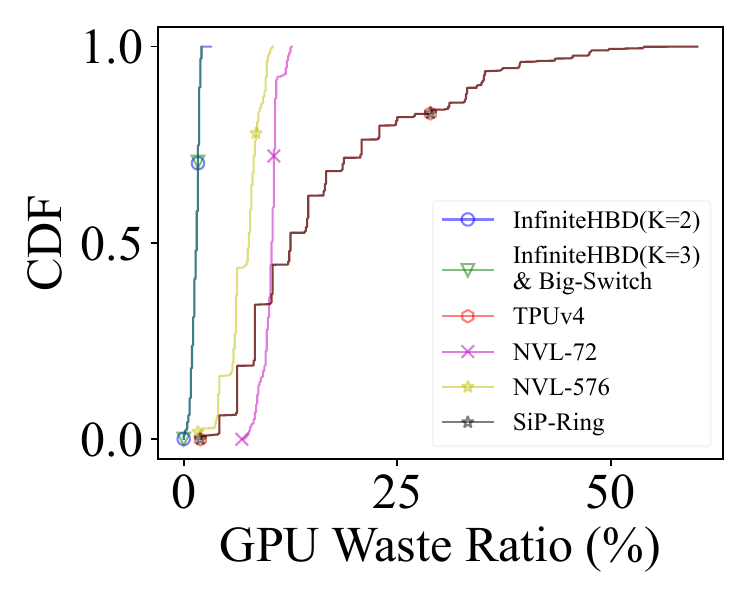}
        \caption{TP-64.}
        \label{fig:simulation:waste-cdf:tp64-gr4}
    \end{subfigure}
    \vspace{-1ex}
    \caption{CDF of GPU waste ratio over production fault trace, 4-GPU node.}
    \label{fig:simulation:waste-cdf:gr4:supple}
\end{figure*}

\begin{figure*}[h!t]
    \centering
    \begin{subfigure}[b]{0.23\linewidth}
        \centering
        \includegraphics[width=\linewidth]{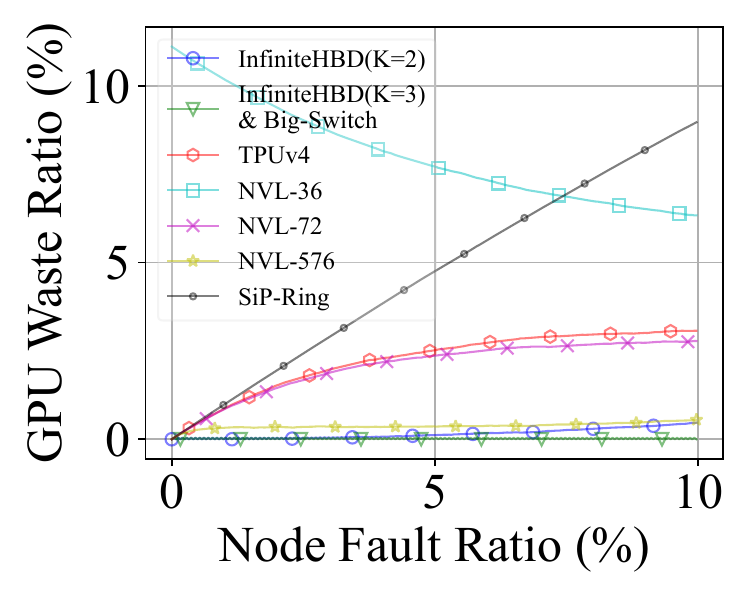}
        \caption{TP-8.}
        \label{fig:simulation:model:wasted:tp8}
    \end{subfigure}
    \hspace{2pt}
    \begin{subfigure}[b]{0.23\linewidth}
        \centering
        \includegraphics[width=\linewidth]{figs/evaluation/fault_model_based/frag_ratio_tp16_gr4.pdf}
        \caption{TP-16.}
        \label{fig:simulation:model:wasted:tp16}
    \end{subfigure}
    \hspace{2pt}
    \begin{subfigure}[b]{0.23\linewidth}
        \centering
        \includegraphics[width=\linewidth]{figs/evaluation/fault_model_based/frag_ratio_tp32_gr4.pdf}
        \caption{TP-32.}
        \label{fig:simulation:model:wasted:tp32}
    \end{subfigure}
    \hspace{2pt}
    \begin{subfigure}[b]{0.23\linewidth}
        \centering
        \includegraphics[width=\linewidth]{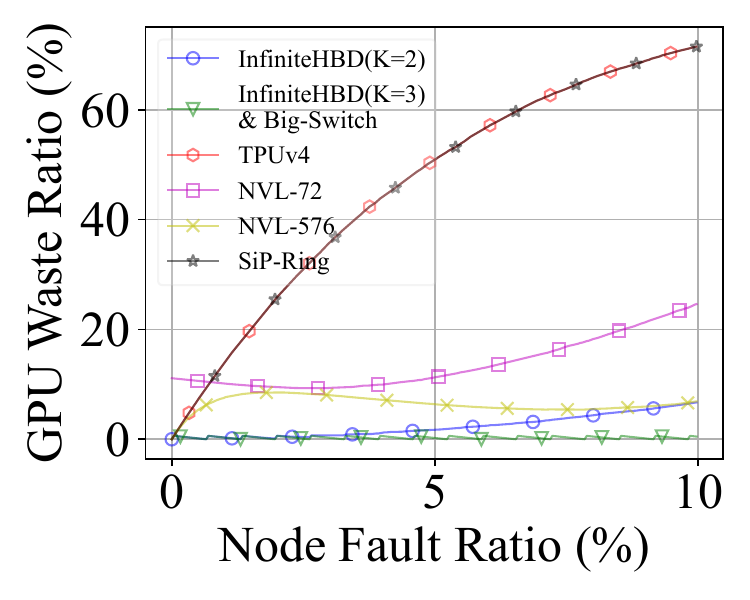}
        \caption{TP-64.}
        \label{fig:simulation:model:wasted:tp64}
    \end{subfigure}
    \vspace{-1ex}
    \caption{GPU waste ratio with different GPU fault ratio, 4-GPU node.}
    \label{fig:simulation:model:wasted-gr4}
\end{figure*}

\begin{figure*}[h!t]
    \centering
    \begin{subfigure}[b]{0.23\linewidth}
        \centering
        \includegraphics[width=\linewidth]{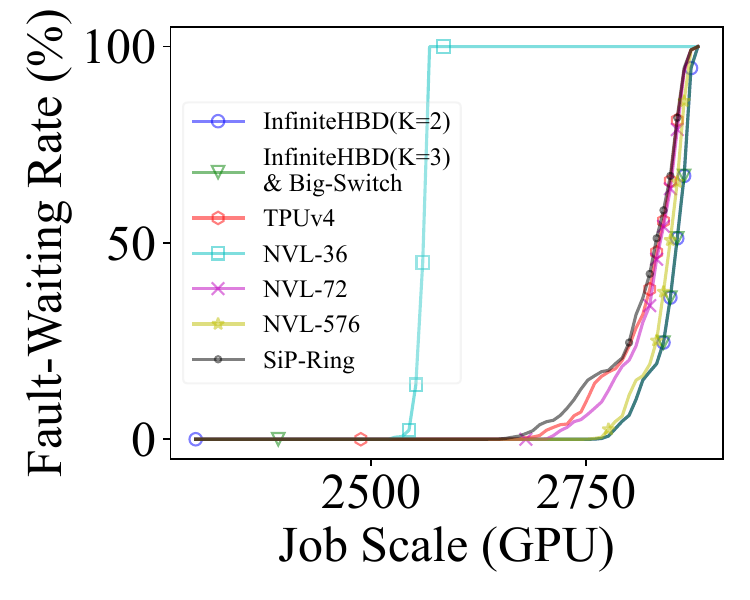}
        \caption{TP-8.}
        \label{fig:simulation:breakdown-duration:tp8-4gpu}
    \end{subfigure}
    \hspace{2pt}
    \begin{subfigure}[b]{0.23\linewidth}
        \centering
        \includegraphics[width=\linewidth]{figs/evaluation/fault_trace_based/breakdown_ratio_tp16_gr4.pdf}
        \caption{TP-16.}
        \label{fig:simulation:breakdown-duration:tp16-4gpu}
    \end{subfigure}
    \hspace{2pt}
    \begin{subfigure}[b]{0.23\linewidth}
        \centering
        \includegraphics[width=\linewidth]{figs/evaluation/fault_trace_based/breakdown_ratio_tp32_gr4.pdf}
        \caption{TP-32.}
        \label{fig:simulation:breakdown-duration:tp32-4gpu}
    \end{subfigure}
    \hspace{2pt}
    \begin{subfigure}[b]{0.23\linewidth}
        \centering
        \includegraphics[width=\linewidth]{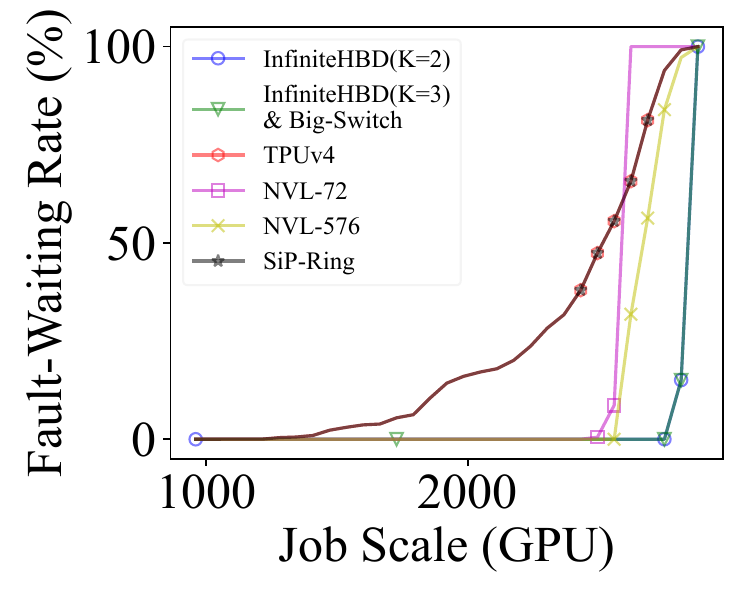}
        \caption{TP-64.}
        \label{fig:simulation:breakdown-duration:tp64-4gpu}
    \end{subfigure}
    \vspace{-1ex}
    \caption{Job fault-waiting duration under different job scales, 4-GPU node.}
    \label{fig:simulation:breakdown-duration-supple}
\end{figure*}

This section presents additional simulation results related to \S\ref{sec:simulation:fault}. \figref{fig:simulation:wasted-trace} shows the variation of the GPU waste ratio over time under the production fault trace. \figref{fig:simulation:waste-cdf:gr4:supple} presents the CDF data for the GPU waste ratio. \figref{fig:simulation:model:wasted-gr4} illustrates the GPU waste ratio for different HBD architectures under various node failure rates, including the results for TP-8 to TP-64. \figref{fig:simulation:breakdown-duration-supple} shows the proportion of job-fault waiting time relative to total time for different job scales. All the aforementioned experiments include results for TP-8, TP-16, TP-32, and TP-64 configurations.

\section{Detailed Cost and Power Consumption Analysis}
\label{appendix:cost}
In this section, \tabref{tab:eval:components} provides a detailed description of the quantity, cost, bandwidth, and power consumption of the interconnect components in various network architectures, including Google TPUv4~\cite{isca2023tpu}, NVIDIA GB200 NVL series~\cite{nvl72}, Alibaba HPN\cite{sigcomm2024hpn}, and \sys{}.

\begin{table*}[h!t] \small
    \centering
    \begin{tabular}{lllll}
    \toprule
    
    \textbf{Component} & \textbf{Quantity} & \textbf{Unit Cost (\$)}  & \textbf{Unit Bandwidth (GBps)} & \textbf{Unit Power (W)} \\

    \midrule
    \multicolumn{5}{c}{\textbf{Google TPUv4\cite{isca2023tpu} with 4096 GPU, bandwidth 300GBps/GPU}} \\
    
    \midrule
    OCS\cite{sigcomm2023lightwave} & 48 & 80000 & 6400 & 108 \\
    DAC Cable\cite{400G_DAC} & 5120 & 63.60 & 50 & 0.1 \\
    Optical Module\cite{400G_OPTICAL_MODULE} & 6144 & 360 & 50 & 12  \\
    Fiber\cite{FIBER}& 6144 & 6.80 & 50 & 0 \\
    
    \midrule
    \multicolumn{5}{c}{\textbf{NVIDIA GB200 NVL-36\cite{SEMIANALYSIS_GB200} with 36 GPU, bandwidth 900GBps/GPU}}\\
    \midrule
    NVLink Switch\cite{SEMIANALYSIS_Power} & 9 & 28000 & 3600 & 275 \\
    DAC Cable\cite{200G_DAC} & 2592 & 35.60 & 25 & 0.1 \\
    
    \midrule
    \multicolumn{5}{c}{\textbf{NVIDIA GB200 NVL-72\cite{nvl72}\cite{SEMIANALYSIS_GB200} with 72 GPU, bandwidth 900GBps/GPU}}\\
    \midrule
    NVLink Switch\cite{SEMIANALYSIS_Power} & 18 & 28000 & 3600 & 275 \\
    DAC Cable\cite{200G_DAC} & 5184 & 35.60 & 25 & 0.1 \\
    \midrule
    \multicolumn{5}{c}{\textbf{NVIDIA GB200 NVL-36x2\cite{SEMIANALYSIS_GB200} with 72 GPU, bandwidth 900GBps/GPU}}\\
    \midrule
    NVLink Switch\cite{SEMIANALYSIS_Power} & 36 & 28000 & 3600 &  275\\
    DAC Cable\cite{200G_DAC} & 6480 & 35.60 & 25 & 0.1 \\
    ACC Cable\cite{SEMIANALYSIS_Power} & 162 & 320 & 200 & 2.5 \\

    \midrule
    \multicolumn{5}{c}{\textbf{NVIDIA GB200 NVL-576\cite{SEMIANALYSIS_GB200} with 576 GPU, bandwidth 900GBps/GPU}}\\
    \midrule
    NVLink Switch\cite{SEMIANALYSIS_Power} & 432 & 28000 & 3600 & 275 \\
    DAC Cable\cite{200G_DAC} & 41472 & 35.60 & 25 & 0.1 \\
    Optical Module\cite{OSFPXD} & 4608 & 850 & 200 & 25 \\
    Fiber\cite{FIBER} & 4608 & 6.80 & 200 & 0 \\

    \midrule
    \multicolumn{5}{c}{\textbf{Alibaba HPN\cite{sigcomm2024hpn} with 16320 GPU, bandwidth 50GBps/GPU}}\\
    \midrule
    EPS\cite{51.2T_EPS} & 360 & 14960 & 6400 & 3145 \\
    DAC Cable\cite{200G_DAC} & 32640 & 35.60 & 25 & 0.1\\
    Optical Module\cite{400G_OPTICAL_MODULE} & 28800 & 360 & 50 & 12 \\
    Fiber\cite{FIBER} & 14400 & 6.80 & 50 & 0 \\

    \midrule
    \multicolumn{5}{c}{\textbf{\SYS{}($K=2$)  with 4 GPU, bandwidth 800GBps/GPU}}\\
    \midrule
    DAC Cable\cite{1.6T_DAC}& 4 & 199.60 & 200 & 0.1\\
    OCSTrx & 16 & 600 & 100 & 12 \\
    Fiber\cite{FIBER} & 16 & 6.80 & 100 & 0 \\

    \midrule
    \multicolumn{5}{c}{\textbf{\SYS{}($K=3$)  with 4 GPU, bandwidth 800GBps/GPU}}\\
    \midrule
    DAC Cable\cite{1.6T_DAC} & 2 & 199.60 & 200 & 0.1\\
    OCSTrx & 24 & 600 & 100 & 12 \\
    Fiber\cite{FIBER} & 24 & 6.80 & 100 & 0 \\
    \bottomrule
    \end{tabular}
    \caption{Interconnect cost and power consumption of components used in different network architectures.}
    \label{tab:eval:components}
\end{table*}

\section{AllToAll Communication of \sys{}}
\label{appendix:all2all}

\revised{Nowadays, training MoE models has become one of the primary application scenarios for GPU clusters. For large MoE models such as DeepSeek-v3~\cite{deepseekv3}, both training and inference heavily rely on expert parallelism (EP). As analyzed earlier (\S\ref{sec:background:workload}), while TP can still effectively support MoE model training, for the sake of generality, we have \textbf{theoretically} explored how \sys{} can be utilized to support EP AllToAll communication. Based on the \docs{} Fast Switch Mechanism, we propose a method to enable the Binary Exchange AllToAll algorithm, though this approach has \textbf{not yet been validated on real hardware}.}

\subsection{\docs{} Fast Switch Mechanism} 
\label{appendix:all2all:fast_switch}

\revised{In the original assumption, \sys{} maintains a fixed topology configuration during any collective communication operation, meaning that only pre-activated links are used for communication. These links form a ring topology, which is inefficient for supporting AllToAll communication. This limitation is particularly common in traditional optical interconnect networks, where the end-to-end reconfiguration latency of OCS—composed of both hardware switching latency and control plane latency—is typically on the order of milliseconds, and in some cases even minutes~\cite{missionapollo}, making it impractical to dynamically reconfigure for each communication pair.}

\revised{However, \docs{} adopts a more simplified hardware architecture that significantly reduces the reconfiguration latency compared to centralized OCS-based switches. Specifically, the hardware switching latency of our \docs{} module is only 60–80~\textmu s. Moreover, the control plane latency can be optimized and minimized by preloading Top-Session configurations in its control module and triggering flow switching as needed. As a result, the total end-to-end reconfiguration latency of \docs{} is reduced to approximately 60-80 $\mu s$. We refer to this combined capability as the \textbf{\docs{} Fast Switch} mechanism.}

\revised{The Fast Switch mechanism allows us to fully utilize all backup links instead of being restricted to pre-activated ones. This means that active links can be dynamically configured in real-time based on any communication pair’s requirements, thereby improving the efficiency of AllToAll communication.}

\subsection{Binary Exchange AllToAll Algorithm}
\label{appendix:all2all:binary-exchange}

\revised{Fast Switch enables the system to fully utilize all available links. However, since nodes equipped with \docs{} have a limited radix, the topology cannot achieve a fully connected structure at any arbitrary scale or location, but rather supports only \textbf{sparse interconnects}. Additionally, architectures like \sys{} do not support \textbf{node-level loopback} at arbitrary scales.}  

\revised{Due to these limitations, AllToAll algorithms based on fully connected topologies (such as those used in NCCL~\cite{NCCL-ALLTOALL}) are not suitable for sparse interconnect environments. Similarly, the lack of node-level loopback makes algorithms that rely on sequential communication infeasible, even if they are inherently sparse. For instance, algorithms like Bruck~\cite{bruck} and Pairwise Exchange~\cite{pairwise-exchange}, which depend on node-level loopback communication, cannot be directly applied to \sys{}.}  

\revised{To better accommodate the interconnect characteristics of \sys{}, we adopt the \textbf{Binary Exchange AllToAll} algorithm. In Binary Exchange, a node $ i $ only communicates with node $ i \oplus 2^k $. For example, in a communication group of 8 nodes, node 0 only exchanges data with nodes 1, 2, and 4 in different rounds. This communication pattern ensures sparsity and eliminates the need for node-level loopback, making it highly compatible with the topology of \sys{}.}

\revised{The detailed procedure of Binary Exchange is shown in \algref{alg:binary-exchange}. Suppose the AllToAll group consists of $ p $ nodes, and each node initially stores $ p\cdot m $ units of data. Each node maintains the following two variables:}

\begin{itemize}[itemsep=2pt,topsep=0pt,parsep=0pt, leftmargin=2ex]
    \item \revised{$Msg$: Stores all data currently held by the node, including both its original data and any received data.}
    \item \revised{$Commset$: Records the set of nodes that have already exchanged data (either directly or through intermediate nodes). Initially, this set contains only the node’s own index.} 
\end{itemize}

\revised{The algorithm runs for $ \log_2 p $ rounds. In the $ k $-th step ($ 1 \leq k \leq \log_2 p $), node $ i $ exchanges data with node $ r = i \oplus 2^{\log_2 p - k} $:}  
\begin{enumerate} [itemsep=2pt,topsep=0pt,parsep=0pt, leftmargin=2ex]
    \item \revised{Node $ i $ sends the following data fragment to node $ r $:  
   $$
   Msg[Commset][r \mathbin{\&} 2^{\log_2 p - k} : r \mathbin{\&} 2^{\log_2 p - k} + 2^{\log_2 p - k}]
   $$  
   and receives data from node $ r $. The transmitted data size per round is $ \frac{p\cdot m}{2} $.}
   \item \revised{The communication set is updated:  
   $$
   i.Commset = i.Commset \cup r.Commset
   $$  }
\end{enumerate}

\revised{The total execution time of Binary Exchange is:  
$$
T = \sum_{i=1}^{\log_2 p} \left( t_s + t_w \frac{pm}{2} \right) = t_s \log_2 p + \frac{1}{2}t_wm p \log_2 p = O(p \log_2 p)
$$
where $ t_s $ is the transmission setup time and $ t_w $ is the per-unit data transfer time.}  

\revised{This algorithm achieves a significantly lower communication complexity, making it well-suited for sparse interconnect architectures without node-level loopback, such as \sys{}. Compared to the $ O(p^2) $ complexity of ring-AllToAll when Fast Switch is not used, Binary Exchange reduces the complexity to $ O(p \log_2 p) $, greatly improving communication efficiency.}

\begin{algorithm}[!h]
\small
\caption{\revised{Binary-Exchange-AllToAll}}
\label{alg:binary-exchange}
\SetAlgoNlRelativeSize{-1}
\SetAlgoNlRelativeSize{1}
 \KwIn{Group size $p$, Node index $i$, Initial message $m_{init}$}
 \KwOut{AllToAll personalized data}
 Initialize $Msg[i]=m_{init}$, $Commset=\{i\}$\;
\For{ $k$ in $1,...,log_2p$}
{
    $r=i \oplus 2^{log_2p-k}$\;
    $m_{send}=\{Msg[m][n]|m\in Commset, n \in [r \mathbin{\&} 2^{\log_2 p - k}, r \mathbin{\&} 2^{\log_2 p - k} + 2^{\log_2 p - k})\}$\;
    Node $i$ sends message $m_{send}$ to node $r$\;
    Node $i$ receives message $m_{recv}$ from node $r$\;
    Update $Msg$ with $m_{recv}$\;
    $Commset=Commset\cup r.Commset$\;   
}
\KwRet {$\{Msg[m][i]|m \in Commset\}$}
\end{algorithm}

\begin{figure*}[!tp]
    \centering
    \includegraphics[width=\linewidth]{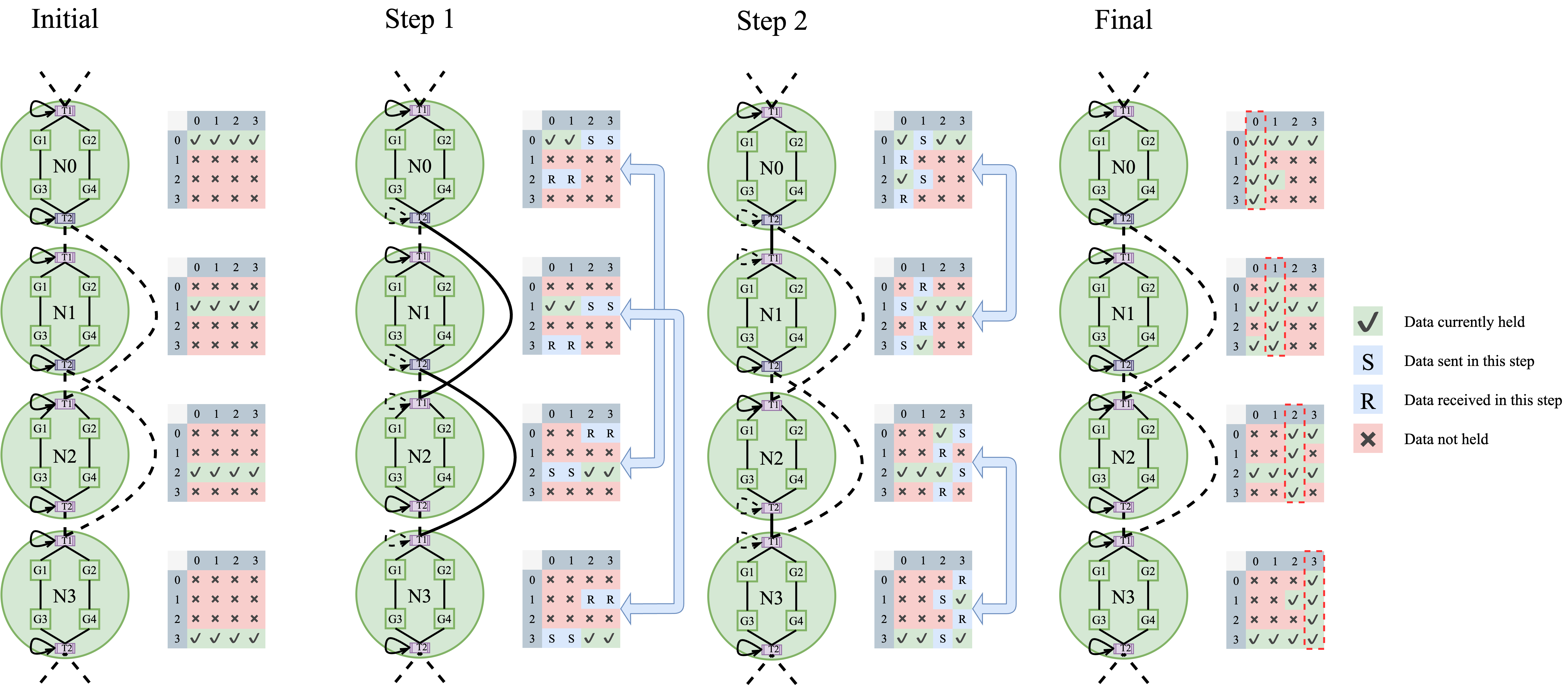}
    \vspace{-5ex}
    \caption{\revised{Illustration of Binary Exchange AllToAll algorithm under TP4+EP4 configuration and \sys{} topology for AllToAll.}}
    \label{fig:all2all}
\end{figure*}

\subsection{\sys{} Topology for AllToAll}
\label{appendix:all2all:topo}

\revised{To support dynamically scalable Binary Exchange AllToAll, we propose a novel topology based on \sys{}.}  

\revised{The topology retains the one-dimensional arrangement of nodes in \sys{} but modifies the wiring pattern. Since in Binary Exchange AllToAll, node $i$ only communicates with node $i \oplus 2^k$, the distance between any two communicating nodes satisfies:  
\[
\Delta = i - (i \oplus 2^k) \in \{+2^k, -2^k\}
\]  
Based on this, instead of connecting each node to neighbors at distances $\pm 1, 2, ..., K$ as in the original K-Hop Ring topology, we rewire nodes to connect at distances $\pm 1, 2, 4, ..., 2^{K-1}$, aligning with the Binary Exchange AllToAll communication pattern.}  

\revised{This topology supports 2D parallelism of \textbf{hybrid TP+EP} in \sys{}, where TP is the inner-layer parallelism and EP is the outer-layer parallelism. 
\figref{fig:all2all} illustrates a TP4+EP4 configuration based on a 4-GPU Node setup, where TP4 is applied within each 4-GPU node, and EP4 is applied across four such nodes.
The figure shows the steps of the Binary Exchange AllToAll algorithm used for EP4 communication: in Step 1, Node 0 exchanges data with Node 2, and Node 1 with Node 3; in Step 2, Node 0 exchanges data with Node 1, and Node 2 with Node 3.}

\revised{However, this topology fundamentally emphasizes a trade-off between AllToAll performance and fault resilience, thereby introducing the following significant challenges:}

\begin{itemize}[itemsep=2pt,topsep=0pt,parsep=0pt, leftmargin=2ex]
    \item \revised{\textbf{Additional GPU forwarding overhead}: Since interconnects are not GPU-level, some traffic must be forwarded through intermediate GPUs, introducing extra communication overhead.}
    
    \item \revised{\textbf{Coupling between TP and EP}: In 4-GPU Node, the number of OCSTrx bundles is limited, allowing each node to connect at most to $\pm 1, 2, 4, 8$ distant nodes. As a result, the TP group size affects the EP group size, constrained by: $TP_{\text{size}} \times EP_{\text{size}} \leq 64 $. This limits general applicability in some scenarios. The 8-GPU Node alleviates this issue, as it supports up to 8 OCSTrx bundles, enabling connections at distances $\pm 1, 2, 4, 8, 16, 32, 64, 128$, relaxing the constraint to:  $TP_{\text{size}} \times EP_{\text{size}} \leq 2048$, making it suitable for most scenarios.}
   
    \item \revised{\textbf{High scheduling complexity}: Node failures affect not only their own availability but also reduce bandwidth on backup links. Scheduling algorithms must simultaneously optimize GPU utilization and inter-node bandwidth, increasing system complexity.}   
\end{itemize} 
\end{appendices}

\end{document}